%% file: VecNetCode-final-v2.tex
\documentclass[letter]{IEEEtran}

\usepackage{caption}

\usepackage[utf8]{inputenc}
\usepackage{graphicx}
\usepackage{amssymb}
\usepackage[cmex10]{amsmath}
\usepackage{color}
\usepackage{theorem}
\usepackage{hyperref}
\usepackage{cite}
\usepackage{url}
\usepackage{breakurl}
\usepackage[english]{babel}
\usepackage[english = american]{csquotes}
\usepackage{mathtools}
\usepackage[ruled,vlined,titlenumbered]{algorithm2e}

\makeatletter
\newcommand{\removelatexerror}{\let\@latex@error\@gobble}
\makeatother

\IEEEoverridecommandlockouts
\MakeOuterQuote{"}

\definecolor{institut_color_orig}{rgb}{0.8,0.3,0.4}
\definecolor{blue_aw}{rgb}{0.3,0.3,1.0}

\usepackage{pgf}
\usepackage{tikz}
\usetikzlibrary{matrix,chains,positioning,arrows,calc,decorations,plotmarks,patterns, fit,backgrounds}
\usepackage{pgfplots}
\usetikzlibrary{external}                       % load externalization library
\usepackage{tkz-graph}
\pgfplotsset{compat=1.3}
\tikzstyle{help lines}=[black!20,dashed]

\pgfplotscreateplotcyclelist{mylist}{red,blue,black,yellow,brown}

\pgfdeclarepatternformonly{my north east lines}{\pgfqpoint{-1pt}{-1pt}}{\pgfqpoint{8pt}{8pt}}{\pgfqpoint{6pt}{6pt}}%
{
  \pgfsetlinewidth{0.4pt}
  \pgfpathmoveto{\pgfqpoint{0pt}{0pt}}
  \pgfpathlineto{\pgfqpoint{6.1pt}{6.1pt}}
  \pgfusepath{stroke}
}
\definecolor{light_gray}{rgb}{0.6,0.6,0.6}
\definecolor{awgray}{rgb}{0.7,0.7,0.7}
\definecolor{awgray_dark}{rgb} {0.4,0.4,0.4}
\tikzset{
    >=stealth',
    mycircle/.style={circle, draw=gray, very thick, text width=.1em, minimum height=1.5em, text centered},
    mycircle_small/.style={circle,draw=awgray_dark,fill = awgray_dark, inner sep=0,minimum size=.6em},
    mycircle_small_black/.style={circle,draw=black,fill = black, inner sep=0,minimum size=.6em},
    mybox/.style={rectangle,rounded corners,draw=black, thick,text width=1em,minimum height=4em,minimum width=4em,text centered},
    mybox_small/.style={rectangle,rounded corners,draw=black, thick,text width=1em,minimum height=2em,minimum width=2em,text centered},
    mybox_vec/.style={rectangle,rounded corners,draw=black, thick,text width=1em,minimum height=0.7em, minimum width=4em,text centered},
    mybox_vec_short/.style={rectangle,rounded corners,draw=black, thick,text width=1em,minimum height=0.7em, minimum width=2em,text centered},
    pil/.style={->, thick, shorten <=2pt, shorten >=2pt,},
}

\input{my_defs-v1.tex}

\begin{document}

\title{\huge Vector Network Coding Based on Subspace Codes Outperforms Scalar Linear Network Coding}
\author{\IEEEauthorblockN{Tuvi Etzion, \textit{Fellow, IEEE}, Antonia Wachter-Zeh, \textit{Member, IEEE}}\\
%\IEEEauthorblockA{Computer Science Department\\ Technion---Israel Institute of Technology, Haifa, Israel\\
%\texttt{\{antonia, etzion\}@cs.technion.ac.il}
\thanks{
T. Etzion is with the Computer Science Department, Technion---Israel Institute of Technology, Haifa, Israel (e-mail: etzion@cs.technion.ac.il).

A. Wachter-Zeh is with the Institute for Communications Engineering, Technical University of Munich, Munich, Germany (e-mail: {antonia.wachter-zeh@tum.de}).	
This work was partly done while A. Wachter-Zeh was with the  Computer Science Department, Technion---Israel Institute of Technology, Haifa, Israel.
	
T. Etzion was supported in part by the Israeli Science Foundation (ISF), Jerusalem, Israel, under Grant 10/12.

At the Technion, A. Wachter-Zeh was supported by the European Union’s Horizon 2020 research and innovation programme under the Marie Sklodowska-Curie grant agreement No 655109 and at TUM by the Technical University of Munich---Institute for Advanced Study, funded by the German Excellence Initiative and European Union Seventh Framework Programme under Grant Agreement No.~291763 and by the German Research Foundation (Deutsche Forschungsgemeinschaft,
DFG) unter Grant No. WA3907/1-1.

Parts of this work have been presented at the \emph{IEEE International Symposium on Information Theory 2016}, Barcelona, Spain, July 2016, \cite{EtzionWachterzeh-ISIT2016}.

}
}

\maketitle
\begin{abstract}
This paper considers vector network coding solutions based on rank-metric codes and subspace codes.
The main result of this paper is that vector solutions can significantly reduce the required alphabet size
compared to the optimal scalar linear solution for the same multicast network.
The multicast networks considered in this paper have one source with $h$ messages,
and the vector solution is over a field of size $q$ with vectors of length~$t$.
For a given network, let the smallest field size for which the network
has a scalar linear solution be $q_s$, then the \emph{gap} in the alphabet size
between the vector solution and the scalar linear solution is defined to be $q_s-q^t$.
In this contribution, the achieved gap is $q^{(h-2)t^2/h + o(t)}$
for any $q \geq 2$ and any even $h \geq 4$.
If $h \geq 5$ is odd, then the
achieved gap of the alphabet  size is $q^{(h-3)t^2/(h-1) + o(t)}$.
Previously, only a gap of size size one had been shown
for networks with a very large number of messages.
These results imply the same gap of the alphabet size between the optimal scalar linear and some scalar \emph{nonlinear} network coding solution
for multicast networks. For three messages, we also show an advantage of vector network coding,
while for two messages the problem remains open.
Several networks are considered, all of them are generalizations and modifications
of the well-known combination networks.
The vector network codes that are used as solutions for those
networks are based on subspace codes, particularly
subspace codes obtained from rank-metric codes. Some of these codes form a new family
of subspace codes, which poses a new {research} problem.
\end{abstract}

%\newpage
%\vspace{-2ex}
\section{Introduction}\label
{sec:intro}

%\PARstart{N}
N{etwork} coding has been attracting increasing attention over the last fifteen years.
%Network coding has been attracting increasing attention over the last fifteen years.
The trigger for this interest was Ahlswede \emph{et al.}'s seminal
paper~\cite{Ahlswede_NetworkInformationFlow_2000}, which revealed that
network coding increases the throughput compared to simple routing.
This gain is achieved since in network coding, the nodes are allowed
to forward a function of their received packets, while in routing
packets can only be forwarded.
The network coding problem can be formulated as follows: given a network with
a set of sources (where each one has a set of messages), for each edge find a function of the packets received at
the starting node of the edge, such that each receiver can recover all its requested information from its received packets.
Such an assignment of a function to each edge is called a \emph{solution} for the network.
Therefore, the received packets on an edge can be expressed as functions of the messages of the sources.
If these functions are linear,
we obtain a \emph{linear network coding solution}, else we speak about a \emph{nonlinear solution}.
In linear network coding, each linear function on an edge
consists of coding coefficients for each incoming packet.
%on an edge can be seen as a coding coefficient
%for each of the incoming packets and each incoming packet is multiplied by its coding coefficient.
If the coding coefficients and the packets are scalars,
it is called a \emph{scalar network coding solution}.
A network which has a solution is called a \emph{solvable} network.
Throughout this paper, we use the short-hand terms \emph{scalar linear solution} and \emph{scalar nonlinear solution}.
In~\cite{KoetterMedard-AlgebraicApproachNetworkCoding_Journal}, K\"{o}tter and M\'{e}dard provided
an algebraic formulation for the linear network coding problem
and its scalar solvability.

Vector network coding as part of \emph{fractional network coding} was mentioned in~\cite{CannonsDoughertyFreilingZeger-2006}.
A solution of the network is called an $(s,t)$ \emph{fractional vector network coding solution},
if the edges transmit vectors of length $t$, but the message vectors
are of length $s \leq t$. The case $s=t=1$ corresponds to a scalar solution.
Ebrahimi and Fragouli~\cite{EbrahimiFragouli-AlgebraicAlgosVectorNetworkCoding}
have extended the algebraic approach from~\cite{KoetterMedard-AlgebraicApproachNetworkCoding_Journal} to
\emph{vector network coding}.
Here, all packets are vectors of length~$t$ and the coding coefficients are matrices.
A set of $t \times t$ coding matrices for which all receivers can recover their
requested information, is called a \emph{vector network coding solution} (henceforth, it will be called \emph{vector solution}).
Notice that vector operations imply linearity over vectors; therefore, a vector solution is always a (vector) \emph{linear} solution.
In terms of the achievable rate, vector network coding outperforms scalar
linear network coding~\cite{MedardKargerEffrosKargerHo_2003,DoughertyFreilingZeger-NetworksMatroidsNonShannon_2007}.
In \cite{MedardKargerEffrosKargerHo_2003}, an example of a network which is not scalar linear
solvable but is solvable by vector routing was shown.
A generalization was given in \cite{DasRai-2016}: Das and Rai proved that
there exists a network with a vector linear solution of dimension~$m$ but
with no vector linear solution over any finite field when the dimension is less than $m$.
Furthermore, in~\cite{DoughertyFreilingZeger-InsufficiencyOfLinearCodingInNetworkInformationFlow_2005}
it was shown that not every solvable network has a vector solution.
In particular, it is shown that there exists a network which has a nonlinear solution
yet has no vector solutions over any finite field and any vector dimension.
The hardness of finding a capacity achieving
vector solution for a general instance of the network coding problem is proven
in~\cite{LangbergSprintson-HardnessApproxmatingNetworkCodingCapacity}.
Recently, in \cite{ConnellyZeger-2016-1}, Connelly and Zeger have shown that in some cases,
there exist networks with scalar linear solutions over some commutative ring but not over the field of the same size.
In~\cite{ConnellyZeger-2016-2}, it was proven that there exist networks with scalar linear solutions
over some non-commutative ring, but not over any commutative ring, and such networks have
vector linear solutions over a field but no scalar linear solutions over any field.

The \emph{alphabet size} of the network coding solution is an important parameter that directly
influences the complexity of the calculations at the network nodes and edges.
In practical applications, it is usually desired to work over small binary
finite fields in order to represent bits and bytes.
The minimum required alphabet size has been a subject for comprehensive research
throughout the last fifteen years of research on network coding, cf. Langberg's and Sprintson's tutorial~\cite{LangbergSprintson-RecentResultsComplexityNEC}.

Throughout this paper, we consider only (single source) \emph{multicast networks}.
A recent survey on the fundamental properties of network coding for multicast
networks can be found in~\cite{Fragouli-Sojanin-NetworkCodingMulticast_2015}.
A multicast network has one source with $h$ messages and all receivers want to receive all the $h$ messages simultaneously.
Notice that such a network is equivalent to a network with $h$ sources, where each source transmits one message.
Li, Yeung, Cai proved that any multicast network is solvable over every sufficiently large field \cite{LiYeungCai-LinearNetworkCoding-2003}.
Jaggi \emph{et al.}~\cite{JaggiSanders-PolyTimeAlgoMulticastNetworkCode} have shown
a deterministic algorithm for finding a scalar linear solution for multicast networks
whose field size is the least prime power that is at least
the number of receivers~$N$.
The algorithm from~\cite{LangbergSprintsonBruck-NetworkCodingComputational} reduces the complexity to find such a solution for the network.
Lehmann and Lehmann~\cite{LehmanLehman-ComplexityNetworkInfoFlow} proved
that there are networks where the linear and nonlinear scalar
solutions both require a field of size in the order of $\sqrt{N}$.
In general, finding the minimum required field size of a (linear or nonlinear) scalar network code for a certain multicast network is NP-complete~\cite{LehmanLehman-ComplexityNetworkInfoFlow}.

For a given network, a vector solution can be transformed into a nonlinear scalar solution.
Dougherty \emph{et al.} have investigated in~\cite{DoughertyFreilingZeger-LinearitySolvabilityMulticastNetworks_2004} several differences
between scalar linear and scalar nonlinear solutions.
For example, they showed that for any integer $h \geq 3$, there exists a multicast network with $h$
messages that has a binary nonlinear solution, but no binary \emph{linear} solution.
Further, they also showed that a network that has a scalar solution over some alphabet might not
have a scalar solution over a larger alphabet (which might not be a finite field).
In~\cite{SunYinLiLong-MulticastNCandFielsSize}, two multicast networks were given:
one of which is solvable over the finite field $\F_7$ but not over $\F_8$,
and one of which is solvable over $\F_{16}$ but not over $\F_{17}$.
They provided the so-called \emph{Swirl network} which is linearly solvable over~$\F_5$ but not
over any $\F_{2^m}$, where $2^m \leq h+2$ and $2^{m}-1$ is a Mersenne prime.

In a scalar linear solution, each coding coefficient can be chosen from $q_s$ values (if the solution is over a field of size~$q_s$).
In vector network coding over a field of size $q$ and dimension~$t$,
each coefficient is a $t \times t$ matrix and can be chosen from~$q^{t^2}$
possibilities. Therefore,
{vector network coding} offers more freedom in choosing the coding
coefficients than does scalar linear coding for equivalent alphabet sizes, and a smaller alphabet size might be
achievable~\cite{EbrahimiFragouli-AlgebraicAlgosVectorNetworkCoding}.

This paper considers a widely studied family of networks, the combination networks, and several generalizations and
modifications of them.
We analyze the scalar linear and vector solutions of these networks. The proposed vector
solutions are based on rank-metric codes and subspace codes.
%at most cubic complexity in $t$.
The main result of our paper is that for several of the analyzed networks, our vector
solutions significantly reduce the required \emph{alphabet size}.
In one subfamily of these networks,
the scalar linear solution requires a field size $q_s = q^{(h-2)t^2/h + o(t)}$, for even $h \geq 4$,
where $h$ denotes the number of messages,
while we provide a vector solution of field size $q$ and dimension~$t$.
Such a vector solution has the same alphabet size as a scalar solution of field size $q^t$, and we denote $q_v \triangleq q^t$.
Therefore, the achieved gap between the alphabet size of the optimal scalar linear solution and our vector
solution is $q^{(h-2)t^2/h + o(t)}$ for any even $h \geq 4$.
Notice that throughout this paper whenever we refer to such a \emph{gap}, we mean the difference
between the \emph{smallest} field (alphabet) size for which a scalar linear solution exists
and the \emph{smallest} alphabet size for which we can construct a vector solution, i.e., the gap is $q_s-q_v$.
For odd $h\geq 5$, the
achieved gap is $q^{(h-3)t^2/(h-1) + o(t)}$.
To our knowledge, so far Sun \emph{et al.}~\cite{SunYangLongLi-MulticastNetworksVectorLinearCoding,SunYangLongLi-MulticastNetworksVectorLinearCoding-arxiv}
has been the only work which presents such a gap but only of size one.
We improve significantly upon~\cite{SunYangLongLi-MulticastNetworksVectorLinearCoding}.
Further, the network of~\cite{SunYangLongLi-MulticastNetworksVectorLinearCoding} has a large number of messages, whereas
our results are based on simple networks and hold for any number of messages $h \geq 4$.
For three messages and certain parameters, we provide a network
in which the vector solution outperforms the optimal scalar linear solution but with a smaller gap.
For two messages, the problem is open, and we conjecture that there
is no advantage in the alphabet size when vector network coding is used.
Finally, in the framework of~\cite{EbrahimiFragouli-AlgebraicAlgosVectorNetworkCoding},
the coding matrices for the vector solutions have to be commutative, while in our solutions they
are not necessarily commutative.

The rest of this paper is structured as follows.
Section~\ref{sec:prelim} provides notations and definitions of finite fields, network coding,
and discusses rank-metric codes and subspaces codes.
Section~\ref{sec:comb-network} considers
combination networks and derives their optimal scalar linear solutions and vector
solutions based on rank-metric codes. Although our vector solutions do not
provide an improvement in terms of the alphabet size for the unmodified
combination networks, it helps to understand the principle of our vector solutions.
Section~\ref{sec:overview-networks} gives an overview of the
generalized combination networks for which a reduction in the alphabet
size will be shown.
In Section~\ref{sec:comb-extra-link-removed-receivers}, we present scalar linear
and vector solutions for some generalized combination networks with additional direct links
from the source to each receiver. Further,
the nodes in these networks are connected via parallel links.
The vector solutions for these networks are based on rank-metric codes, and the required alphabet size is significantly reduced.
In particular, the largest achieved gap between scalar and vector network coding is
$q^{(h-2)t^2/h + o(t)}$ for any even integer $h \geq 4$.
In Section~\ref{sec:subspaces}, we show that the constructions, which are
based on rank-metric codes, can be seen as constructions based on subspace codes.
Moreover, using subspace codes, some results can be improved.
Section~\ref{sec:analyzing}
analyzes and compares more generalized combination networks and the achieved gap
in the alphabet size of the optimal scalar linear solutions and our vector solutions.
The vector solutions for these networks are based on subspaces codes.
In particular, a gap between scalar and vector network coding of size
$q^{(h-3)t^2/(h-1) + o(t)}$ is obtained for any odd integer $h \geq 5$ messages.
A network where the vector solution also improves the alphabet size compared to
the optimal scalar solution for $h=3$ messages is shown in Section~\ref{sec:network-3-messages}.
This network, as well as other similar networks, poses a new interesting problem related to subspace codes.
Finally, concluding remarks are given in Section~\ref{sec:conclusion}, and
several open problems for future research are outlined.

\section{Preliminaries}\label{sec:prelim}
\subsection{Finite Fields}
Let $q$ be a power of a prime, $\Fq$ denote the finite field of order $q$, and $\Fqm$ its extension field of order~$q^m$.
We use~$\Fq^{m \times n}$ for the set of all $m\times n$ matrices over $\Fq$.
Let $\mathbf I_{s}$ denote the~$s \times s$ identity matrix and
$\0_s$ the $s\times s$ all-zero matrix.
The triple $\codelinearArb{n,k,d}_{q}$ denotes a linear code over $\Fq$ of length~$n$, dimension~$k$, and minimum Hamming distance $d$.
Let $\Fq[x]$ denote the ring of polynomials with coefficients in $\Fq$.

\subsection{Network Coding}
A \emph{network} will be modeled as a finite directed acyclic multigraph,
with a set of source nodes and a set
$\myset{R} = \{R_1,\dots, R_N\}$ of $N$ receivers.
The sources have sets of disjoint messages, which are symbols {or vectors} over a given finite field.
To unify the description, we refer to \emph{packets} for both cases, symbols and vectors.
Each receiver demands (requests) a subset of the $h$ messages.
Each edge in the network has unit capacity, and it carries a packet which is either a
symbol from $\F_{q_s}$ (in scalar network coding) or a vector of length $t$ over $\Fq$ (in vector network coding).
Note that the assumption of unit capacity does not restrict the considered networks,
since edges of larger capacity can be represented by multiple parallel edges of unit capacity.
The incoming and outgoing edges of a node~$V$ are denoted by $In(V)$ and $Out(V)$, respectively.

A \emph{network code} is a set of functions of the packets on the edges
of the network. For a source $S$, the edges in $Out(S)$ carry functions
of the messages of $S$. For any vertex $V$, which is not a source or a receiver,
the edges in $Out(V)$ carry functions of the packets on the edges of $In(V)$.
The network code is called \emph{linear} if all the functions are linear, and nonlinear otherwise.
A network code is a \emph{solution} for the network if each receiver can reconstruct
its requested messages from the packets on its incoming edges.
Such a network is called \emph{solvable}.
The network code is called a \emph{scalar network code} if the packets are scalars
from $\F_{q_s}$ and thus, each edge carries a scalar from~$\F_{q_s}$.
The network code is called a \emph{vector network code} if the packets
are vectors and each edge carries a vector of length $t$
with entries from a field~$\F_q$.

A \emph{(single source) multicast network} is a network with exactly one source,
where all receivers demand all the $h$ messages simultaneously.
This is possible if there exist $h$ edge disjoint paths from the source to each receiver.
This is equivalent to saying that the \emph{min-cut}
between the source and each receiver is at least $h$,
where the min-cut is the minimum number of edges that have
to be deleted to disconnect all sources from the receiver.
The network coding \emph{max-flow/min-cut theorem} for multicast networks states that the maximum number of messages transmitted from the source to each receiver is equal to the smallest min-cut between
the source and any receiver~\cite{Ahlswede_NetworkInformationFlow_2000}.
In the sequel, we will only write \emph{network} instead of \emph{multicast network} since this paper considers only multicast networks.

To formalize this description, let $x_1,\dots,x_h$ denote the $h$ source messages for scalar linear network coding.
Each edge $E$ in $Out(V)$, for any given vertex $V$, builds a linear combination of the symbols obtained from $In(V)$.
The coefficient vector of such a linear combination is called the \emph{local coding vector}.
Clearly, from all the functions on the paths leading to $V$,
the packet of $E$ can be written as a linear combination of the~$h$ messages.
The coefficients of this linear combination are called the \emph{global coding vector}.
Each receiver finally obtains several linear combinations of the~$h$ message symbols (its global coding vectors).
Thus, any receiver $R_j$, $j \in \{1,\dots,N\}$, has to solve the following linear system of equations:
\begin{equation*}
\begin{pmatrix}
y_1\\
\vdots\\
y_k
\end{pmatrix}
=
\Mat{A}_j \cdot
\begin{pmatrix}
x_1\\
\vdots\\
x_{h}
\end{pmatrix},
\end{equation*}
where $k = |In(R_j)|$, and $\Mat{A}_j$ is a $k \times h$ \emph{transfer matrix} which contains the global
coding vectors on the edges of $In(R_j)$,
and the symbols on $In(R_j)$ are $\{y_1,\dots,y_k\}$.
In scalar linear network coding, we therefore want to find edge coefficients
such that the matrix $\Mat{A}_j$ has full rank for every $j=1,\dots,N$.
These coefficients should have field size $q_s$ as small as possible.

In \emph{vector network coding}, the edges transmit vectors, and therefore, the coding coefficients at each node are matrices.
A \emph{vector solution} is said to have dimension $t$ over a field of size~$q$
if all these vectors are over a field of size $q$ and have length~$t$, where $q_v = q^t$.
Let $\x_1,\dots,\x_h$ denote $h$ source messages which are now vectors of length~$t$.
Then, any receiver $R_j$ has to solve the following linear system of equations:
\begin{equation*}
\begin{pmatrix}
\y_1\\
\vdots\\
\y_k
\end{pmatrix}
=
\Mat{A}_j \cdot
\begin{pmatrix}
\x_1\\
\vdots\\
\x_{h}
\end{pmatrix},
\end{equation*}
where $k = |In(R_j)|$, and $\Mat{A}_j$ is a $(kt) \times (ht)$ \emph{transfer matrix}
which contains the global coding vectors (vectors whose entries are matrices) on the edges of $In(R_j)$,
and the vectors on $In(R_j)$ are $\{\y_1,\dots,\y_k\}$.
In vector network coding, we therefore want to find edge coefficients (which are matrices)
such that the matrix $\Mat{A}_j$ has full rank for each $j=1,\dots,N$ and such that $q_v=q^t$ is minimized.

When we want to compare scalar linear and vector network coding,
the solutions are equivalent with respect to the alphabet size when $q_s = q_v$.

\subsection{Rank-Metric Codes}

Codewords of rank-metric codes will be used in some of our constructions as coding coefficients for vector solutions.
Let $\rk(\A)$ be the rank of a matrix $\A\in \Fq^{m \times n}$.
The \emph{rank distance} between $\A , \B \in \Fq^{m\times n}$ is defined by
$d_{\fontmetric{R}}(\A,\B)\triangleq  \rk(\A-\B)$.
A linear $\codelinearRank{m \times n,k,\delta}$ {rank-metric code} \mycode{C} is a $k$-dimensional subspace of~$\Fq^{m \times n}$. It consists of $q^k$ matrices of size $m \times n$ over $\Fq$ with minimum rank distance:
\begin{equation*}
\delta \triangleq
%\min_{\substack{{\A,\B} \in \mycode{C}\\ \A \neq \B}}
%\big\lbrace d_{\fontmetric{R}}(\A,\B) =
%\rk(\A-\B) \big\rbrace
%=
\min_{\substack{{\A} \in \mycode{C}, \A \neq \0}}
\big\lbrace \rk(\A) \big\rbrace.
\end{equation*}
The Singleton-like upper bound for rank-metric
codes~\cite{Delsarte_1978,Gabidulin_TheoryOfCodes_1985,Roth_RankCodes_1991}
implies that for any $\codelinearRank{m \times n,k,\delta}$ code, we have that $k \leq \max\{m,n\}(\min\{n,m\}-\delta+1)$.
Codes which attain this bound with equality are known
for all feasible parameters~\cite{Delsarte_1978,Gabidulin_TheoryOfCodes_1985,Roth_RankCodes_1991}.
They are called \emph{maximum rank distance} (MRD) codes
and denoted by $\MRDlinq{m\times n, \delta}$.

Let $\Mat{C}$ be the \emph{companion matrix} of a primitive polynomial of degree $t$ over $\F_q$.
The set of matrices $\mycode{D}_t =\{ \0_t , \I_t , \Mat{C} ,\Mat{C}^2 , \ldots , \Mat{C}^{q^t-2} \}$
forms an $\MRDlinq{t\times t, t}$ code of~$q^t$ pairwise \emph{commutative} matrices (see~\cite{Roth_RankCodes_1991,Lusina2003Maximum}).
\begin{theorem}\label{thm:code-square}
Let
%\begin{equation*}
$\mycode{D}_t \triangleq \{ \0_t , \I_t , \Mat{C} ,\Mat{C}^2 , \ldots , \Mat{C}^{q^t-2} \}$,
%\end{equation*}
where $\Mat{C}$ is a companion matrix:
\begin{equation*}
\Mat{C} =
\begin{pmatrix}
0 & 1 & 0 & \dots & 0 & 0\\
0 & 0 & 1 & \dots & 0 & 0\\
\vdots\\
0 & 0 & 0 & \dots & 0 & 1\\
-p_0 & -p_1 & -p_2 & \dots & -p_{t-2}& -p_{t-1},
\end{pmatrix},
\end{equation*}
and $p(x) = p_0 + p_1 x + \dots + p_{t-2}x^{t-2} + p_{t-1}x^{t-1} + x^t \in \Fq[x]$ is a primitive polynomial.
Then, $\mycode{D}_t$ is an $\MRDlinq{t\times t, t}$ code.
\end{theorem}

The set of matrices $\mycode{D}_t$ and the finite field $\F_{q^t}$ are isomorphic. In particular, if $\alpha$ is a root of the primitive polynomial $p(x)$, then $\alpha^i \cdot \alpha^j$ can be implemented by multiplying the corresponding two matrices $\C^i \cdot \C^j = \C^{i+j}$, where $\C$ is the companion matrix based on $p(x)$.
Similarly, $\alpha^i + \alpha^j$ corresponds to $\C^i + \C^j$.
{Notice that $\C^0 = \C^{q^t-1} = \I_t$. This
property will be needed later in our constructions.}
These matrices are very useful when we design a vector network code for the combination networks (see Section~\ref{sec:comb-network}).

Moreover, to prove that any network (multicast or non-multicast)
has a vector solution of dimension $t$ over~$\F_q$ if there exists a scalar linear solution
over $\F_{q^t}$, we can simply replace any coefficient $\alpha^s$ by $\C^s$.
Due to the isomorphism, addition and multiplication can be done in matrix representation as well.
Further, the matrices of the code $\mycode{D}_t$
are useful in encoding and decoding used in the network. Instead of computing
in the field $\F_{q^t}$, we can use the related matrices of the code to obtain the vector solution
and translate it to the scalar solution only at the receivers.
%We note that for large $t$ this translation can be costly.

\subsection{Subspace Codes}
\label{subsec:prilim-subspaces}
In our constructions of vector network codes, the global coding vector consists of
$h$ matrices from $\F_q^{t \times t}$.
These~$h$ matrices can be appended together to form a $t \times (ht)$ matrix
which is a basis for a subspace of $\F_q^{ht}$ whose dimension is at most $t$. In the sequel of the paper,
we will see that if we use a set of subspaces spanned by such matrices with additional properties,
then they can be used as our coding coefficients.
%The set of such matrices
%is a \emph{subspace code}.

Let $\langle \Mat{A} \rangle$ denote the space spanned by the rows of a matrix~$\Mat{A}$.
Similarly, for $k$ vectors $\vec{a}_1, \dots, \vec{a}_k$, let
$\langle\vec{a}_1, \dots, \vec{a}_k\rangle$ denote the space spanned by these vectors.
The \emph{Grassmannian} of dimension $k$
is denoted by $\Grassm{n,k}$ and
is the set of all subspaces of $\Fq^n$ of dimension $k \leq n$.
The cardinality of $\Grassm{n,k}$ is the well-known $q$-binomial (also known as Gaussian coefficient):
\begin{equation*}
\big|\Grassm{n,k}\big|=\quadbinom{n}{k} \triangleq \prod\limits_{i=0}^{k-1} \frac{q^n-q^i}{q^k-q^i},
\end{equation*}
where
$q^{k(n-k)}\leq \quadbinom{n}{k} < 4 q^{k(n-k)}$.
The set of all subspaces of $\F_q^n$ is called the \emph{projective space of order $n$}
and is denoted by $\cP_q(n)$, i.e., $\cP_q(n) = \bigcup_{k=0}^n \Grassm{n,k}$.
For two subspaces $\myspace{U},\myspace{V} \in \cP_q(n)$, let $\myspace{U}+\myspace{V}$
denote the smallest subspace containing the union of $\myspace{U} \in \cP_q(n)$ and $\myspace{V} \in \cP_q(n)$.
The \emph{subspace distance} between $\myspace{U} \in \cP_q(n)$ and $\myspace{V} \in \cP_q(n)$ is defined by
$\Subspacedist{\myspace{U},\myspace{V}}
\triangleq2 \dim(\myspace{U}+\myspace{V})-\dim(\myspace{U})-\dim(\myspace{V})$.
A \emph{subspace code} is a set of subspaces; if the subspaces in the subspace code have the same dimension,
then the code is called a \emph{constant dimension code} or a \emph{Grassmannian code}. These codes were considered
for error-correction in random linear network coding~\cite{koetter_kschischang}. Bounds on the sizes of such codes and properties which
are relevant for our discussion can be found in~\cite{Etzion2009ErrorCorrecting,EtzionSilberstein-CodesDesignsRelLiftedMRD-2013,Etzion2011ErrorCorrecting,EtzionGorlaRavagnaniWachterzeh_Ferrers_2014}.
Let $A_q(n,k,d)$ denote the maximum cardinality of a constant
dimension code in $\Grassm{n,k}$ with minimum subspace distance~$d$.
The following bounds can be found in~\cite{koetter_kschischang,EtzionSilberstein-CodesDesignsRelLiftedMRD-2013}.
\begin{theorem}
\label{thm:bound_sub_codes}
For $\delta >1$,
\begin{equation*}
q^{(n-k)(k-\delta +1)}\leq A_q(n,k,2\delta) \leq 2q^{(n-k)(k-\delta +1)},
\end{equation*}
and for $\delta =1$
\begin{equation*}
q^{(n-k)k}\leq A_q(n,k,2)=\quadbinom{n}{k} \leq 4q^{(n-k)k}.
\end{equation*}
\end{theorem}

\section{The Combination Networks}
\label{sec:comb-network}

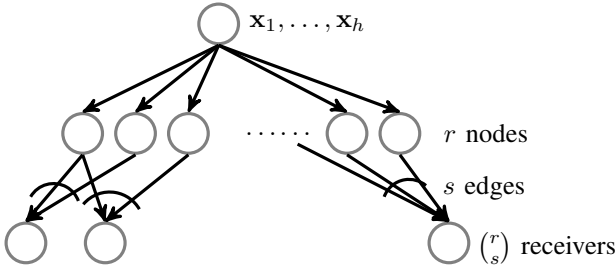
\begin{figure}[htb]
\input{comb-network.tex}
\caption{The $\mathcal{N}_{h,r,s}$ combination network: it has an edge from the source to each of the $r$ nodes
in the middle layer. Each of the $\binom{r}{s}$ receivers is connected to a unique set of $s$ middle-layer nodes
and demands all of the messages $\vec{x}_1 ,\ldots , \vec{x}_h$.
}
\label{fig:comb-net}
\end{figure}

The $\mathcal{N}_{h,r,s}$ combination network (where $s \geq h$) is shown in Fig.~\ref{fig:comb-net} (see also~\cite{RiisAhlswede-ProblemsNetworkCodingECC}).
The network has three layers: the first layer consists of a source with $h$ messages.
The source transmits $r$ packets to the $r$ nodes of the middle layer. Any~$s$ nodes in the middle layer are
connected to a receiver, and each one of the $\binom{r}{s}$ receivers demands all the $h$ messages.
For vector coding, the messages $\vec{x}_1,\dots, \vec{x}_h$ are vectors of length~$t$, and
for scalar coding, the messages are scalars, denoted by $x_1, \dots, x_h$.
In a combination network, the local and the global coding vectors are the same, and therefore, we will not distinguish between the two.
Next, we consider the case where $s=h$.

\subsection{Scalar Solution}

The $\mathcal{N}_{h,r,h}$ combination network has a scalar linear solution
of field size $q_s$ if and only if an $\codelinearArb{r,h,d=r-h+1}_{q_s}$ MDS code exists~\cite{RiisAhlswede-ProblemsNetworkCodingECC}.
From the known theory on MDS codes~\cite[pp.~317--331]{MacWilliamsSloane_TheTheoryOfErrorCorrecting_1988} it is known that such a code
exists for every prime power $q_s$ such that
$q_s \geq r-1$. The only exceptions are when $h \in \{3,q_s-1\}$ and $q_s$ is a power of two, where
such a code exists for every prime power $q_s$ such that ${q_s \geq r-2}$~\cite[p.~328]{MacWilliamsSloane_TheTheoryOfErrorCorrecting_1988}.
The symbols transmitted to and from each node in the middle layer form together
a codeword of the MDS code (encoded from the~$h$ message symbols),
and each receiver obtains~$h$ symbols. Each receiver can correct $r-h$ erasures and therefore can reconstruct the~$h$ message symbols.

\begin{corollary}
\label{cor:decoding-compl-scalar-comb-gen}
If $h=3$ and $r-2$ is a power of two, let $q^*=r-2$, else let $q^*=r-1$.
For the $\mathcal{N}_{h,r,h}$ combination network, a scalar linear solution of field size $q_s$ exists if and only if $q_s \geq q^*$.
%The best known decoding complexity is in $\min\{\OCompl{r\log^2 r},\OCompl{h^{2.37}}\}$ over a field of size~$q^*$.
\end{corollary}
%Thus, for $r \in \{q^t+1, q^t+2\}$, the decoding complexity is at least $\OCompl{q^t}$ over a field of size at least $q^t$ for each receiver.

%\vspace{-2ex}
\subsection{Vector Solution}
\label{sec:vector-coding-combination}

In the sequel, we present a vector solution based on MRD codes for the $\mathcal{N}_{h,r,h}$ combination network.
The case $h=2$ was solved similarly in~\cite{SunYangLongLi-MulticastNetworksVectorLinearCoding}.
%These results can be used to generalize the $N$-butterfly network to multiple inputs as well.
%We will need the following theorem whose proof can be found in the appendix.
Our construction uses the isomorphism between the field $\mathbb{F}_{q^t}$ and the powers of the companion matrix, i.e., the set $\mycode{D}_t$ (Theorem~\ref{thm:code-square}).
This isomorphism leads to the following theorem.
\begin{theorem}
Let $\alpha \in \F_{q^t}$ be a root of the primitive polynomial $p(x)$ of $\F_{q^t}$.
Let $\Mat{M} \in \F_{q^t}^{\ell \times \ell}$ be an arbitrary matrix.
Define the block matrix $\Mat{M}^\prime \in \Fq^{(\ell t) \times (\ell t)}$ by replacing each entry of $\Mat{M}$ as follows: if $M_{ij} = \alpha^{k}$, $k \in \{0, \dots, q^t-2\}$, replace it by $\C^k$ for all $i,j =1,\dots, \ell$, where $\C$ denotes the companion matrix based on $p(x)$. If $M_{ij} =0$, then replace it by $\0_{t}$.

Any set of $\lambda$ linearly independent columns of $\Mat{M}$
is linearly independent over $\F_{q^t}$ if and only if
the columns of the related $\lambda$ blocks of columns in~$\Mat{M}^\prime$ are linearly independent over $\Fq$.
\end{theorem}
\begin{IEEEproof}
%Within one block of columns, the $t$ columns are clearly all linearly independent since the companion matrix and its powers have full rank.
%Columns between different blocks are linearly independent over $\Fq$ since
%they were linearly independent over $\F_{q^t}$ in $\Mat{M}$.
Denote by $\Mat{M}_{\lambda}$ a full-rank $\ell \times \ell$ submatrix of $\Mat{M}$.
Define a corresponding $(\ell t) \times (\ell t)$ matrix $\Mat{M}^\prime_{\lambda}$ over $\Fq$ by replacing $\alpha^{k}$ by $\C^k$ and $0$ by $\0_t$.
Clearly, the determinant of $\Mat{M}_{\lambda}$ is a function of its entries $M_{\lambda,ij}$, $i,j=1,\dots,\ell$.
The determinant of $\Mat{M}^\prime_{\lambda}$ has the same form as $\det(\Mat{M}_{\lambda})$,
with the only difference that each $M_{\lambda,ij} = \alpha^k$ is replaced
by $\C^{k}$ (and if $M_{\lambda,ij} = 0$, it is replaced by~$\0_t$).
Thus, $\det(\Mat{M}^\prime_{\lambda})$ is non-zero if and only if
$\det(\Mat{M}_{\lambda})$ is non-zero and the difference and the product of any two distinct
matrices $\C^i$ and $\C^j$ (where at least one of them is non-zero) has full rank.
The second property is true since the powers of the companion matrix form
a full-rank MRD code, see Theorem~\ref{thm:code-square}, and hence, the statement
follows.
\end{IEEEproof}

The following corollary considers block Vandermonde matrices which will be used for our vector solution.
Recall that $\I_t = \C^{q^t-1}\in \mycode{D}_t$.
\begin{corollary}\label{cor:q-vand-block-matrix}
Let $\mycode{D}_t$ % = \{ \0_t , \I_t , \Mat{C} ,\Mat{C}^2 , \ldots , \Mat{C}^{q^t-2} \}$
be the $\MRDlinq{t \times t,t}$ code defined by the companion matrix $\C$
(Theorem~\ref{thm:code-square}).
Let $\Mat{C}_i$, $i=1,\dots,h$, be distinct codewords of~$\mycode{D}_t$.
Define the following $(ht) \times (ht)$ block matrix:
\begin{equation*}
\Mat{M} =
\begin{pmatrix}
\I_t & \Mat{C}_1 & \Mat{C}_1^2 & \dots & \Mat{C}_1^{{h-1}}\\
\I_t & \Mat{C}_2 & \Mat{C}_2^2 & \dots & \Mat{C}_2^{{h-1}}\\
\vdots & \vdots& \vdots & \ddots & \vdots\\
\I_t & \Mat{C}_{h} & \Mat{C}_{h}^2 & \dots & \Mat{C}_{h}^{{h-1}}\\
\end{pmatrix}.
\end{equation*}
Then, any  $(h t) \times (\ell t)$ submatrix consisting of $h \ell$ blocks of
\textbf{consecutive columns} has full rank $\ell t$, for any $\ell=1,\dots,h$.
\end{corollary}
Note that the blocks of rows do not have to consecutive, but a block has to be included with all its $t$ rows in the submatrix.

Based on this corollary, we can now provide a vector network code.

\begin{construction}
\label{constr:comb-network-gen}
Let $\mycode{D}_t =\{\Mat{C}_1, \Mat{C}_2,\dots, \Mat{C}_{q^t}\}$ be the $\MRDlinq{t \times t,t}$ code defined by the companion matrix~$\C$ (Theorem~\ref{thm:code-square}) and let $r \leq q^t+1$. Consider the $\mathcal{N}_{h,r,h}$ combination network with message vectors $\vec{x}_1, \dots, \vec{x}_h$.
One node from the middle layer receives and transmits $\vec{y}_{r} = \vec{x}_h$ and
the other $r-1$ nodes of the middle layer receive and transmit
$\vec{y}_{i} =
\begin{pmatrix}
\I_t \ \C_i \ \C_i^2 \ \dots \ \C_i^{h-1}
\end{pmatrix}
\cdot
\begin{pmatrix}
\vec{x}_1 \
\vec{x}_2 \
\dots \
\vec{x}_h
\end{pmatrix}^T
\in \Fq^t$,
%\end{equation*}
for $i=1,\dots,r-1$.
%such that $\Mat{C}_i\neq\Mat{C}_j$ for $i\neq j$ and $\Mat{C}_i \in \mycode{C}$, $\forall i\in \{2,\dots,r\}$.
\end{construction}
The matrices $\I_t,\C_i, \C_i^2,\dots,\C_i^{h-1}$, $i=1,\dots,r-1$, are the coding coefficients of the incoming and outgoing edges of node $i$ in the middle layer.
\begin{theorem}\label{thm:solution-combination}
Construction~\ref{constr:comb-network-gen} provides a vector linear solution of field size $q$ and
dimension $t$ to the $\mathcal{N}_{h,q^t+1,h}$ combination network, i.e., $\vec{x}_1, \dots, \vec{x}_h$ can be reconstructed at all receivers.
\end{theorem}
\begin{IEEEproof}
Each receiver obtains $h$ vectors and has to solve one of the following two systems of linear equations:
\begin{equation*}
\begin{pmatrix*}
\vec{y}_{i_1}\\
\vdots\\
\vec{y}_{i_{h-1}}\\
\vec{y}_{i_h}
\end{pmatrix*}
=
\begin{pmatrix}
\I_t & \Mat{C}_{i_1} & \Mat{C}_{i_1}^2 & \dots & \Mat{C}_{i_1}^{{h-1}}\\
\I_t & \Mat{C}_{i_2} & \Mat{C}_{i_2}^2 & \dots & \Mat{C}_{i_2}^{{h-1}}\\
\vdots & \vdots& \vdots & \ddots & \vdots\\
\I_t & \Mat{C}_{i_h} & \Mat{C}_{i_h}^2 & \dots & \Mat{C}_{i_h}^{{h-1}}\\
\end{pmatrix}.
\begin{pmatrix}
\vec{x}_1\\
\vec{x}_2\\
\vdots\\
\vec{x}_h
\end{pmatrix}
\end{equation*}
or
\begin{equation*}
\begin{pmatrix*}
\vec{y}_{i_1}\\
\vdots\\
\vec{y}_{i_{h-1}}\\
\vec{y}_1
\end{pmatrix*}
=
\begin{pmatrix}
\I_t & \Mat{C}_{i_1} & \Mat{C}_{i_1}^2 & \dots & \Mat{C}_{i_1}^{{h-1}}\\
%\I & \Mat{C}_{i_2} & \Mat{C}_{i_2}^2 & \dots & \Mat{C}_{i_2}^{{k-1}}\\
\vdots & \vdots& \vdots & \ddots & \vdots\\
\I_t & \Mat{C}_{i_{h-1}} & \Mat{C}_{i_{h-1}}^2 & \dots & \Mat{C}_{i_{h-1}}^{{h-1}}\\
\0_t & \0_t & \0_t & \dots & \I_t
\end{pmatrix}.
\begin{pmatrix}
\vec{x}_1\\
\vec{x}_2\\
\vdots\\
\vec{x}_h
\end{pmatrix},
\end{equation*}
for some distinct $i_1,\dots,i_h \in \{2,\dots,r\}$.
Due to Corollary~\ref{cor:q-vand-block-matrix}, in both cases, the corresponding matrix
has full rank and there is a unique solution for $(\vec{x}_1 \ \vec{x}_2 \ \dots \ \vec{x}_h)$.
\end{IEEEproof}

The decoding at each receiver consists of solving a linear system of equations of size $(ht) \times (ht)$.
The following theorem analyzes the decoding complexity of this vector solution.
Thereby, we use the MRD code from Theorem~\ref{thm:code-square} which is
formed by the companion matrix and its powers.
Thus, the inverse of these matrices can be calculated with less than quadratic complexity.
\begin{theorem}\label{thm:decoding-comb-gen}
For the $\mathcal{N}_{h,q^t+1,h}$ combination network, a vector solution of field size $q$ and dimension $t$ exists. The decoding complexity is in $\OCompl{ht \log^2 h \log^2 t}$ over $\Fq$ for each receiver.
\end{theorem}
\begin{IEEEproof}
From Theorem~\ref{thm:solution-combination} it is clear that such a solution exists.
For the decoding complexity, it remains to prove that inverting the matrix $\Mat{M}$ from
Corollary~\ref{cor:q-vand-block-matrix} can be done with $\OCompl{ht \log^2 h \log^2 t}$ operations over $\Fq$.
Since the submatrices are commutative, the inverse of this block Vandermonde matrix can be
calculated in the same way as the inverse of a usual Vandermonde matrix~\cite{Turner-InverseVandermonde}.
The only difference is that multiplication and addition of elements from $\Fq$ are now replaced
by multiplication and addition of the commutative code matrices.
%The multiplication of $\C_i=\C^i$ and $\C_j=\C^j$ equals $\C^{i+j}$ and is therefore just writing
%down $t$ elements from $\mathbb{F}_{q^t}$, i.e., has complexity $\OCompl{t^2}$ over $\Fq$.
%Clearly, addition of two such matrices costs also $\OCompl{t^2}$ over $\Fq$. Thus, the
%complexity of inverting $\Mat{M}$ costs $t^2$
Due to the isomorphism between $\F_{q^t}$ and $\myset{D}_t$,
this multiplication and addition is equivalent to fast polynomial multiplication and
fast polynomial addition modulo the primitive polynomial $p(x)$, and thus, the
matrix additions and multiplications are in the order of $\OCompl{t \log^2 t}$ over $\Fq$~\cite{Gohberg94fastalgorithms}.
Thus, the complexity of inverting $\Mat{M}$ costs
$t \log^2 t$ times the complexity of inverting an
$h \times h$ Vandermonde matrix, which is $\OCompl{h \log ^2 h}$~\cite{Gohberg94fastalgorithms}.
\end{IEEEproof}

Further, for the $\mathcal{N}_{3,q^t+2,3}$-combination network with three messages
and when~$q^t$ is a power of two, we can use the matrices from Construction~\ref{constr:comb-network-gen} and additionally transmit
$\vec{x}_2=(\0_t \ \I_t \ \0_t) \cdot (\vec{x}_1 \ \vec{x}_2 \ \vec{x}_3)^T$
to obtain a vector solution.

\subsection{Analysis}
\label{subsec:comparison-comb}
Due to the isomorphism of $\F_{q^t}$ and the code $\mycode{D}_t$, both solutions are
equivalent. Hence, in practice, implementing the scalar solution can actually be done by implementing the vector solution.
We can therefore construct a vector linear solution of field size $q$ and dimension $t$ for the $\mathcal{N}_{h,q^t+1,h}$ combination
network, where equivalently a scalar solution from an MDS code exists for $q_s \geq q^t$.
The decoding complexity when implementing the vector solution is in
$\OCompl{ht \log^2 h \log^2 t}$ operations over $\Fq$ for each receiver.

\section{A Family of Generalized Combination Networks}
\label{sec:overview-networks}

%\subsection{Overview}
In this section, we discuss the
generalizations and modifications of the combination networks which are considered in this paper.
We therefore define a generalization of the $\mathcal{N}_{h,r,s}$ combination network, called the
\emph{$\epsilon$-direct links $\ell$-parallel links $\mathcal{N}_{h,r,s}$ network}, in short the $(\epsilon,\ell)$-$\mathcal{N}_{h,r,s}$ network.
All our considered networks (including the unmodified combination networks)
are special cases of the $(\epsilon,\ell)$-$\mathcal{N}_{h,r,s}$ network.

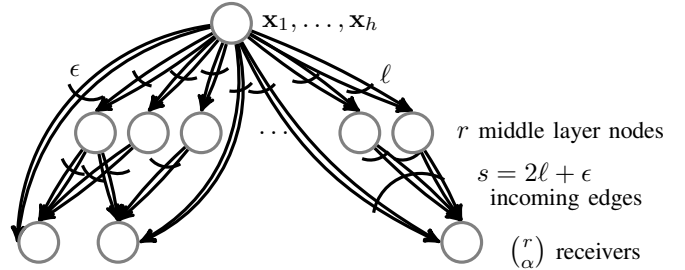
\begin{figure}[htb]
\input{network-ell-tau.tex}
\caption{The $(\epsilon,\ell)$-$\mathcal{N}_{h,r,s}$ network for $\alpha = \frac{s-\epsilon}{\ell}=2$, where each
of the $\binom{r}{\alpha}$ receivers is connected to a unique set of $\alpha$ middle layer nodes and demands
all the messages $\vec{x}_1 ,\ldots , \vec{x}_h$.}
\label{fig:network-ell-tau}
\end{figure}

We start by describing the structure of the
generalized networks and then we will consider the main subfamilies of this family of networks.
An example of an $(\epsilon,\ell)$-$\mathcal{N}_{h,r,s}$ network
is shown in Fig.~\ref{fig:network-ell-tau} and
consists of three layers.
In the first layer, there is a source
with $h$ messages.
%If we refer to a vector solution, these messages are denoted by $\vec{x}_1,\dots, \vec{x}_h$; for a scalar solution, we denote them by $x_1,\dots,x_h$.
In the middle layer, there are $r$ nodes.
The source has $\ell$ parallel links to each node in the middle layer.
From any $\alpha = \frac{s-\epsilon}{\ell}$ nodes
in the middle layer, there are $\ell$ parallel links to one
receiver in the third layer.
Additionally, from the source there are
$\epsilon$ direct parallel links to each one of the $\binom{r}{\alpha}$ receivers in the third layer.
Therefore, each receiver has $s = \alpha\ell+\epsilon$ incoming links.
We will assume some relations between the parameters $h$, $\alpha$, $\epsilon$, and $\ell$
such that the resulting network is interesting and does not have a trivial or no solution (see Theorem~\ref{thm:trivialORno}).

\begin{remark}
Our definition of the generalized combination networks depends on five parameters, $\epsilon$, $\ell$, $h$, $r$, and $s$, where
$\epsilon$ and $\ell$ are parameters which do not appear in the combination networks. The value of $h$ is the number of messages, $r$ is
the number of nodes in the middle layer, and the parameter $s$ is the in-degree of each receiver.
The parameter $\alpha$ denotes the number of edges leaving each middle node.
Hence, one might equivalently define the generalized
combination networks by using $\alpha$ instead of $s$.
\end{remark}

Notice first that the \emph{local} and \emph{global} coding vectors for the
$(\epsilon,\ell)$-$\mathcal{N}_{h,r,s}$ network must be the same
(since only at the source the encoding can make a difference).
Hence, we will not distinguish between them in the sequel.
Second, it should be remarked that if $s > h$, then
the min-cut of the $(\epsilon,\ell)$-$\mathcal{N}_{h,r,s}$ network is larger
than the number of messages~$h$. An equivalent network in which the min-cut
is $h$, which is solved with an alphabet of the same size, can be
constructed as follows: replace the $i$-th receiver $R_i$ by a node $T_i$
from which there are $h$ links to $h$ vertices $P_{ij}$, $1 \leq j \leq h$.
From $P_{ij}$, $1 \leq j \leq h$, there is a link to a new receiver $R_i^\prime$.
Similarly, we can avoid parallel links in the network. Assume there are $\ell$ parallel links
from vertex $U$ to vertex $V$. We can remove these links and add $\ell$ vertices $W_1,W_2,\ldots ,W_\ell$,
such that there exists a link from $U$ to $W_i$, $1 \leq i \leq \ell$, and there exists a link
from each vertex $W_i$, $1 \leq i \leq \ell$, to $V$. Again, the new network is solvable over the same
alphabet as the old network. In the $(\epsilon,\ell)$-$\mathcal{N}_{h,r,s}$ network, this transformation can be done
by simply replacing each node in the middle layer by~$\ell$ nodes.

The following subsections list several instances of the
$(\epsilon,\ell)$-$\mathcal{N}_{h,r,s}$ network.
In the next section, we will analyze the scalar linear solution
and the vector solution for some of these
special $(\epsilon,\ell)$-$\mathcal{N}_{h,r,s}$ networks
and compare them in terms of the alphabet size $q_s$ and $q_v$.

\subsection{The Combination Networks}

The $\mathcal{N}_{h,r,s}$ combination network is clearly a special case of the
$(\epsilon,\ell)$-$\mathcal{N}_{h,r,s}$ network,
where $\epsilon =0$ and $\ell =1$. This network with $h=s$ was already considered
in Section~\ref{sec:comb-network}. We are not aware of any set of parameters
for which a vector solution outperforms the optimal scalar linear solution
in the unmodified combination network with respect to the alphabet size.

On the other hand, we can prove that there are combination networks for which
vector solution cannot outperform scalar linear solution with respect to the
alphabet size. One interesting such an example is the $\mathcal{N}_{2,r,2}$ combination network.
As was pointed out before, such a network has a solution of field size $q_s$ if and only
if an $[r,2,d=r-1]_{q_s}$ MDS code exists~\cite{RiisAhlswede-ProblemsNetworkCodingECC}.
Such a code is also equivalent to a set of $r-2$ pairwise orthogonal Latin squares
of order $q_s$. The largest~$r$ so that such a code and a set can exist is $r=q_s+1$.
For every prime power $q_s$ such a linear code and a set do exist. Hence, there
is no difference between the alphabet size of the scalar linear solution and
the scalar nonlinear solution. Since any vector solution can be translated to a scalar
nonlinear solution over the same alphabet size, it follows that the vector solution
cannot outperform the scalar linear solution with respect to the alphabet size.
Unfortunately, we have no result of this kind for the $\mathcal{N}_{h,r,h}$ combination
network when $h \geq 3$. We elaborate more on this in Section~\ref{sec:conclusion}.

\subsection{The Direct Links Combination Network}\label{subsec:eps-1-network}

\begin{figure}[htb]
\input{comb-network-extra-link.tex}
\caption{The $(1,1)$-$\mathcal{N}_{3,r,4}$ network with $\binom{r}{3}$ receivers.
For the same alphabet size $q_s=q_v=4$, in the optimal scalar linear solution the number of middle layer nodes is $r \leq 42$,
while in the optimal vector solution the number of middle layer nodes is $r \geq 51$ (see Section~\ref{sec:network-3-messages}),
i.e., a vector solution outperforms the optimal scalar linear solution.
}
\label{fig:comb-net-extra}
\end{figure}
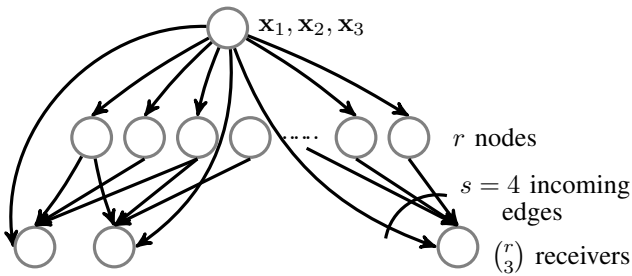

Another interesting family of networks that will be considered
is the combination network with $\epsilon \geq 1$ additional direct links,
i.e., the $(\epsilon,1)$-$\mathcal{N}_{h,r,s}$ network.
This network with $h=s=3$ and $\epsilon=1$ is discussed in Section~\ref{sec:network-3-messages}.
It is the only network with three messages
for which we obtained an advantage of a vector solution compared to the optimal scalar linear solution
with respect to the alphabet size.
Both the optimal scalar linear solution and the given vector solution for this
subfamily of networks motivate some interesting questions on a classic coding
problem and on a new type of subspace code problem, which will be discussed in the sequel.
For $h=s=3$ and $\epsilon=1$ this network is illustrated in Fig.~\ref{fig:comb-net-extra}.

\subsection{The One-Direct Link $\ell$-Parallel Links $\mathcal{N}_{2\ell,r,2\ell+1}$ Network}
\label{subsec:one-ell-network}

\begin{figure}[htb]
\input{comb-network-rate2-ell.tex}
\caption{The $(1,\ell)$-$\mathcal{N}_{2\ell,r,2\ell+1}$ network with $\binom{r}{2}$ receivers.
For the alphabet sizes, we obtain
$q_v=q^t$ and $q_s > q^{t^2/2}$ (see~Section~\ref{subsec:compa12}), where $r=q^{\ell t(t+1)}$,
i.e., when $r=q^{\ell t(t+1)}$, this network has a vector solution with $q_v=q^t$ and scalar
solutions only if $q_s > q^{t^2/2}$.
}
\label{fig:comb-net-rate2-1}
\end{figure}
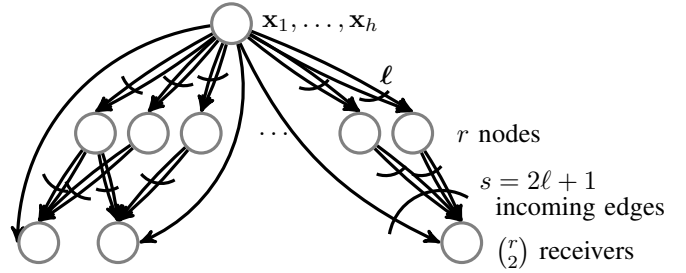

The $(1,\ell)$-$\mathcal{N}_{2\ell,r,2\ell+1}$ network is shown in Fig.~\ref{fig:comb-net-rate2-1}.
It has three layers: a source in the first layer with $h=2\ell$ messages and $r$ nodes in the middle layer, where
there are $\ell$ links from the source to each node in the middle layer. There are $\binom{r}{2}$ receivers, which
form the third layer, where each set of two nodes from the middle layer is connected to
a different receiver. If a node $V$ from the middle layer is connected to a receiver $R$, then there are $\ell$ links from $V$ to $R$.
There is also one direct link from the source to each receiver. The total number of edges entering a receiver is $2\ell +1$.
This is the subfamily with the smallest number of direct links from the
source to the receivers for which our vector solution outperforms
the optimal scalar linear solution.

\subsection{The $(\ell-1)$-Extra Links $\ell$-Parallel Links $\mathcal{N}_{2\ell,r,3\ell -1}$ Network}
\label{subsec:ell-1-ell-network}

\begin{figure}[htb]
\input{comb-network-rate2-ell-plus.tex}
\caption{The $(\ell-1,\ell)$-$\mathcal{N}_{2\ell,r,3\ell-1}$ network with $\binom{r}{2}$ receivers.
For the alphabet sizes, we obtain $q_v=q^t$ and $q_s > q^{t^2(\ell-1)/\ell}$ (see Section~\ref{sec:gen-comb-parallel-extra-links}),
where $r=q^{\ell (\ell-1)t^2+\ell t}$,
i.e., when $r=q^{\ell (\ell-1)t^2+\ell t}$, this network has a vector solution with $q_v=q^t$ and scalar
solutions only if $q_s > q^{t^2(\ell-1)/\ell}$.
}
\label{fig:comb-net-nplus}
\end{figure}
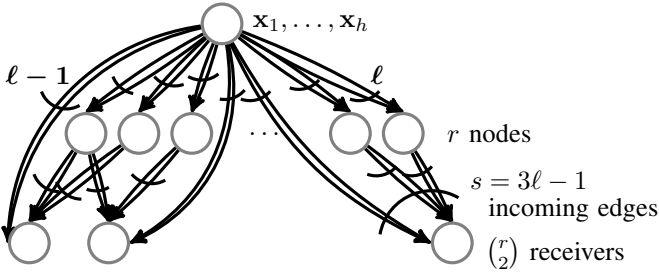

The second subfamily in which vector solutions outperform scalar solutions
is the $(\ell-1,\ell)$-$\mathcal{N}_{2\ell,r, 3\ell -1}$ network, which
is shown in Fig.~\ref{fig:comb-net-nplus}. This is a nontrivial subfamily in which the
number of direct links from the source to the receivers
is the largest one. It has three layers,
with a source carrying $h=2\ell$ messages in the first layer.
In the second layer,
there are $r$ nodes and in the third layer, there are $\binom{r}{2}$ receivers.
This network yields the largest gap in the alphabet size between our vector solution and the optimal scalar solution for an even number $h \geq 4$ of messages.
The intersection with the previous subfamily, the $(1,\ell)$-$\mathcal{N}_{h,r,s}$ network, is when $\ell =2$ and the related network is
the $(1,2)$-$\mathcal{N}_{4,r,5}$ network shown in Fig.~\ref{fig:comb-net-rate2}.

\section{Vector Solutions Using Rank-Metric Codes which Outperform Scalar solutions}
\label{sec:comb-extra-link-removed-receivers}

\subsection{Overview}
In this section, we will consider vector solutions based on
rank-metric codes for some of the generalized combination networks.
We will start with a basic example of the $(1,2)$-$\mathcal{N}_{4,r,5}$ network. The idea of the optimal scalar linear solution
for this network will demonstrate the general idea of all our results
with $h \geq 4$ messages.
The gap between scalar and vector network coding is $q^{t^2/2 + o(t)}$.
The more general network that we consider later is the $(\ell-1,\ell)$-$\mathcal{N}_{2\ell,r,3\ell -1}$ network for which the gap
between scalar and vector network coding
is $q^{(\ell -1) t^2/\ell + o(t)}$ for any $\ell \geq 2$.

\subsection{Scalar Linear Solution for the $(1,2)$-$\mathcal{N}_{4,r,5}$ Network}\label{subsec:scalarsol12}
\begin{figure}[htb]
\input{comb-network-rate2.tex}
\caption{The $(1,2)$-$\mathcal{N}_{4,r,5}$ network with $\binom{r}{2}$ receivers.
For the alphabet sizes, we obtain $q_v=q^t$ and $q_s> q^{t^2/2}$ (see Section~\ref{subsec:scalarsol12}), where $r=q^{2t(t+1)}$,
i.e., when $r=q^{2t(t+1)}$, this network has a vector solution with $q_v=q^t$ and scalar
solutions only if $q_s> q^{t^2/2}$.
	}
\label{fig:comb-net-rate2}
\end{figure}
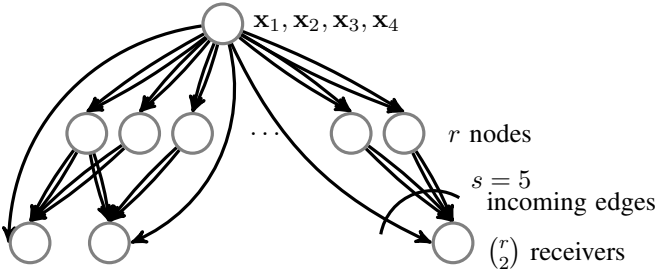

For ease of understanding, the $(1,2)$-$\mathcal{N}_{4,r,5}$ network
is shown in Fig.~\ref{fig:comb-net-rate2}.

\begin{lemma}\label{lem:scalar-lin-rate-two}
There is a scalar linear solution of field size $q_s$ for the $(1,2)$-$\mathcal{N}_{4,r,5}$ network if and only if
$r \leq (q_s^2+1)(q_s^2+q_s+1)$.
\end{lemma}
\begin{IEEEproof}
Let $\Mat{B}$ be a $4 \times (2r)$ matrix, divided into $r$ blocks of two columns,
with the property that any two blocks together have rank at least three.
Each one of the~$r$ nodes in the middle layer of the network transmits two symbols
(from one block) of $(x_1, x_2,x_3,x_4) \cdot \Mat{B}$ (these symbols were
transmitted to the node from the source). Each direct link
transmits a symbol $p_i = \sum_{j=1}^{4} p_{ij} x_j$, for $i=1,\dots,\binom{r}{2}$,
which is chosen such that the related $4 \times 4$-submatrix of $\Mat{B}$ with
the additional vector column $(p_{i1}, p_{i2}, p_{i3},  p_{i4})^T$ has full rank.
Clearly, there is a scalar solution over $\F_{q_s}$ if and only if such a matrix over $\F_{q_s}$ exists.

These blocks are defined to be any $4\times 2$ matrix representations of all two-dimensional subspaces
of $\F_{q_s}^4$. Any two blocks together form a $4\times 4$ matrix of rank at least three (since any two such subspaces are distinct).

From every node in the middle layer, there are two links to the appropriate receivers.
Therefore, we associate each middle node with one block. The number of blocks is at most the
number of distinct two-dimensional subspaces of $\F_{q_s}^4$, i.e., ${r \leq \quadbinoms{4}{2}}$
and therefore, a scalar solution exists if:
\begin{equation*}
r \leq \quadbinoms{4}{2} = (q_s^2+1)(q_s^2+q_s+1).
\end{equation*}

To prove the "only if", we need to show that there is no scheme that provides more blocks.
Assume, one block is a rank-one matrix. Then, all other blocks must have rank two and the space that they span has
to be disjoint to the rank-one block. Therefore, with this scheme there are at most $1 + \quadbinoms{3}{2} < \quadbinoms{4}{2}$ blocks.
Thus, for the largest~$r$ all blocks should have rank two, and taking all two-dimensional subspaces provides the maximum number of blocks.
\end{IEEEproof}

\subsection{Vector Solution for the $(1,2)$-$\mathcal{N}_{4,r,5}$-Network}
\label{subsec:vec-2-rate-k=4}
\begin{construction}
\label{constr:mrd-notfull-rank-comb-2}
Let $\mycode{C}=\{\Mat{C}_1,\Mat{C}_2,\dots,\Mat{C}_{q^{2t^2 + 2t}}\}$ be an $\MRDlinq{2t \times 2t,t}$ code and let $r \leq q^{2t^2 + 2t}$.
Consider the $(1,2)$-$\mathcal{N}_{4,r,5}$ network with message vectors $\vec{x}_1, \vec{x}_2, \vec{x}_3, \vec{x}_4 \in \Fq^t$.
The $i$-th middle node receives and transmits:
\begin{equation*}
\begin{pmatrix}
\vec{y}_{i_1} \\
\vec{y}_{i_2}
\end{pmatrix}
=
\begin{pmatrix}
\I_{2t} & \C_i
\end{pmatrix}
\cdot
\begin{pmatrix}
\vec{x}_1\\
\vec{x}_2\\
\vec{x}_3\\
\vec{x}_4
\end{pmatrix}
\in \Fq^{2t},
\quad
i=1,\dots,r.
\end{equation*}
The direct link from the source which ends in the same receiver as the links from two distinct nodes $i,j \in \{ 1,2,\ldots , r\}$,
of the middle layer transmits the vector $\vec{z}_{ij} = \Mat{P}_{ij} \cdot \begin{pmatrix}
\vec{x}_1,
\vec{x}_2,
\vec{x}_3,
\vec{x}_4
\end{pmatrix}^T\in \Fq^t$,
%\end{equation*}
where the $t \times (4t)$ matrix $\vec{P}_{ij}$ is chosen such that
\begin{equation}\label{eq:mat-constr1}
\rk\begin{pmatrix}
\I_{2t} & \C_i \\
\I_{2t} & \C_j \\
&\hspace{-4ex} \Mat{P}_{ij}
\end{pmatrix} = 4t.
\end{equation}
\end{construction}
Since $$\rk\left(\begin{matrix}
\I_{2t} & \C_i \\
\I_{2t} & \C_j \\
\end{matrix}\right)  =
\rk\left(\begin{matrix}
\I_{2t} & \C_i \\
\0_{2t} & \C_j-\C_i \\
\end{matrix}\right)
\geq 3t,$$
it follows that the $t$ rows of $\Mat{P}_{ij}$ can be chosen such that the overall rank of the matrix from~\eqref{eq:mat-constr1} is $4t$.
\begin{theorem}
\label{thm:solution-extra-rate2}
Construction~\ref{constr:mrd-notfull-rank-comb-2} provides a vector solution of field size $q$
and dimension $t$ to the $(1,2)$-$\mathcal{N}_{4,r,5}$ network for any $ r\leq q^{2t(t+1)}$.
\end{theorem}
\begin{IEEEproof} %[Proof of Theorem~\ref{thm:solution-extra-rate2}]
Each receiver $R_{ij}$ obtains the vectors $\vec{y}_{i_1}, \vec{y}_{i_2},
\vec{y}_{j_1},
\vec{y}_{j_2}$ and the vector ${\vec{z}_{ij}}$ from the direct link.
From these five vectors, the receiver wants to reconstruct the message vectors
$\vec{x}_1,\dots,\vec{x}_4$ by solving the following linear system of equations:
\begin{equation*}
\begin{pmatrix*}
\vec{y}_{i_1}\\
\vec{y}_{i_2}\\
\vec{y}_{j_1}\\
\vec{y}_{j_2}\\
{\vec{z}_{ij}}
\end{pmatrix*}
=
\begin{pmatrix}
\I_{2t} & \C_i\\
\I_{2t} & \C_j\\
& \hspace{-4ex}\Mat{P}_{ij}
\end{pmatrix}
\cdot
\begin{pmatrix}
\vec{x}_1\\
\vec{x}_2\\
\vec{x}_3\\
\vec{x}_4
\end{pmatrix}
\end{equation*}
The choice of $\Mat{P}_{ij}$ from Construction~\ref{constr:mrd-notfull-rank-comb-2} guarantees that this linear system of equations
has a unique solution for $(\vec{x}_1, \vec{x}_2,\vec{x}_3,\vec{x}_4)$.
\end{IEEEproof}

\subsection{Comparison of the Solutions for the $(1,2)$-$\mathcal{N}_{4,r,5}$ Network}\label{subsec:compa12}

For the $(1,2)$-$\mathcal{N}_{4,r,5}$ network, we obtain a significant improvement in the alphabet size
for the vector solution compared to the optimal scalar linear solution.
The alphabet size of the vector solution is $q_v=q^t$, while the field size $q_s$ of
the optimal scalar linear solution has to satisfy $r \leq (q_s^2+1)(q_s^2+q_s+1)$.
Since $r$ can be chosen to be $q^{2t^2+2t}$, the size of the gap is $q^{t^2/2+o(t)}$.

The same gap in the alphabet size can be obtained for the
$(1,\ell)$-$\mathcal{N}_{2\ell,r,2\ell+1}$ network, for any $\ell \geq 2$,
by using the same approach with an $\MRDlinq{\ell t \times \ell t,(\ell-1)t}$ code.

\subsection{Solutions for the $(\ell-1,\ell)$-$\mathcal{N}_{2\ell,r,3\ell-1}$ Network}
\label{sec:gen-comb-parallel-extra-links}

To improve the gap on the alphabet size, i.e., to show that the advantage of our vector solution
compared to the optimal scalar linear solution is even more significant, we consider in this subsection
the $(\ell-1,\ell)$-$\mathcal{N}_{2\ell,r,3\ell-1}$ network from Subsection~\ref{subsec:ell-1-ell-network}.

Let us first consider the optimal scalar linear solution for this network.
\begin{lemma}\label{lem:scalar-sol-nplus}
There is a scalar linear solution of field size $q_s$ for the $(\ell-1,\ell)$-$\mathcal{N}_{2\ell,r,3\ell-1}$ network
with $2\ell$ messages, where $\ell \geq 2$, if and only if $r \leq \quadbinoms{2 \ell}{\ell}$.
\end{lemma}
\begin{IEEEproof}
Let $\Mat{B}$ be a $(2\ell) \times (\ell r)$ matrix, divided into $r$ disjoint blocks of $\ell$ columns,
with the property that any two blocks together have rank at least $\ell+1$.
Each of the~$r$ nodes in the middle layer of the network transmits~$\ell$ symbols
(from one block) of $(x_1, \dots, x_{2\ell}) \cdot \Mat{B}$ (these symbols were
transmitted to the node from the source). Each direct link
transmits a symbol $p_i = \sum_{j=1}^{2\ell} p_{ij} x_j$, for $i=1,\dots,\binom{r}{2}$,
which is chosen such that the related $(2\ell) \times (2\ell)$ submatrix of $\Mat{B}$ with
the additional vector column $(p_{i1}, p_{i2}, p_{i3},  p_{i4})^T$ has full rank.
Clearly, there is a scalar solution over $\F_{q_s}$ if and only if such a matrix over $\F_{q_s}$ exists.

Thus, each of these $r$ blocks forms an $\ell$-dimensional subspace of $\F_{q_s}^{2\ell}$ such
that any two such $\ell$-dimensional subspaces intersect in a subspace of dimension at most $(\ell-1)$.
Hence, the subspace distance between two such subspaces is at least two,
and all $\ell$-dimensional subspaces
of $\Grassms{2 \ell , \ell}$ can be used. The size of $\Grassms{2 \ell , \ell}$  is $\quadbinoms{2 \ell}{\ell}$
which completes the proof.
\end{IEEEproof}

Let us now consider the vector solution.
\begin{construction}
\label{constr:mrd-notfull-rank-comb-2-gen2}
Let $\mycode{C}=\{\Mat{C}_1,\Mat{C}_2,\dots,\Mat{C}_{q^{\ell(\ell-1) t^2 + \ell t}}\}$ be an
$\MRDlinq{\ell t \times \ell t,t}$ code and let $r$ be any integer
such that $r \leq q^{\ell(\ell-1) t^2 + \ell t}$.
Consider the $(\ell-1,\ell)$-$\mathcal{N}_{2\ell,r,3\ell-1}$ network with
message vectors $\vec{x}_1, \dots, \vec{x}_{2\ell} \in \Fq^t$, where $\ell \geq 2$.
The $i$-th middle node receives and transmits:
\begin{equation*}
\begin{pmatrix}
\vec{y}_{i_1} \\
\vec{y}_{i_{2}}
\end{pmatrix}
=
\begin{pmatrix}
\I_{\ell t} & \C_i
\end{pmatrix}
\cdot
\begin{pmatrix}
\vec{x}_1\\
\vdots\\
\vec{x}_{2\ell}
\end{pmatrix}
\in \Fq^{\ell t},
\quad
i=1,\dots,r.
\end{equation*}
The $\ell-1$ direct links from the source, which end at the same receiver as
the links from two distinct nodes $i,j \in \{ 1,2,\ldots , r\}$
of the middle layer, transmit the vectors $\vec{z}_{ijs} = \Mat{P}_{ijs} \cdot \begin{pmatrix}
\vec{x}_1,
\dots, \vec{x}_{2\ell}
\end{pmatrix}^T\in \Fq^t$,
for $s=1,\dots,\ell-1$,
where the $t \times (2\ell t)$ matrices $\vec{P}_{ijs}$ are chosen such that
\begin{equation}\label{eq:matrix-nplus-network}
\rk\begin{pmatrix}
\I_{\ell t} & \C_i \\
\I_{\ell t} & \C_j \\
&\hspace{-4ex} \Mat{P}_{ij1}\\
&\hspace{-4ex} \vdots\\
&\hspace{-4ex} \Mat{P}_{ij(\ell-1)}
\end{pmatrix} = 2\ell t.
\end{equation}
\end{construction}
By the rank distance of $\mycode{C}$ we have that
$\rk\left(\begin{smallmatrix}
\I_{\ell t} & \C_i \\
\I_{\ell t} & \C_j \\
\end{smallmatrix}\right) \geq \ell t + t = (\ell+1)t$, and hence the $(\ell-1)t$
rows of the matrices~$\Mat{P}_{ijs}$ can be chosen such that the overall rank
of the matrix from \eqref{eq:matrix-nplus-network} is~$2\ell t$.

The following result is an immediate consequence of this construction.
\begin{corollary}
\label{thm:solution-extra-rate2-gen2}
Construction~\ref{constr:mrd-notfull-rank-comb-2-gen2} provides a vector solution of field size $q$
and dimension $t$ to the $(\ell-1,\ell)$-$\mathcal{N}_{2\ell,r,3\ell-1}$ network
for any $r \leq q^{\ell(\ell-1) t^2 + \ell t}$ with $2\ell$ messages for each $\ell \geq 2$.
\end{corollary}

For $r = q^{\ell (\ell -1) t^2 +\ell t}$, we have that the field size $q_s$ for any
scalar linear solution
has to satisfy $r= q^{\ell (\ell -1) t^2 +\ell t} \leq \quadbinoms{2 \ell}{\ell} < 4 q_s^{\ell^2}$ (see Section~\ref{subsec:prilim-subspaces}) and therefore the size of the gap between the scalar and vector network coding solutions is
$q^{(\ell-1)t^2/\ell + o(t)}$, for any $\ell \geq 2$.
Further, the size of the gap tends to $q^{t^2 + o(t)}$ for large $\ell$.

\section{Vector Solutions Using Subspace Codes}
\label{sec:subspaces}

In this section, we will improve our vector solutions by using subspace codes.
In the sequel it will be explained how larger gaps, as a function of $t$, between the alphabet size
of the vector solution and the scalar linear solution can be obtained by using large subspace codes.
However, the leading term in the exponent of the alphabet size in the gap will not change, i.e.,
\emph{asymptotically}, there is no improvement compared to the subspace codes based on rank-metric codes.
Moreover, the description with rank-metric codes is simpler and easier to analyze.
%%However, vector solutions using subspace codes reduce the alphabet size in most of the generalizations of the combination networks compared to the alphabet size using rank-metric
%codes by a constant factor, even if not asymptotically.
However, by using larger subspace codes, we may be able to find vector solutions which outperform
the optimal scalar solution (with respect to the alphabet size) in cases where this cannot be done using rank-metric codes.
Finally, as we will see in Section~\ref{sec:network-3-messages}, there are important networks on which
such improvements occur.
Note that we have already used subspace codes for the scalar solutions
of the generalized combination networks (see, e.g., Lemma~\ref{lem:scalar-lin-rate-two} and Lemma~\ref{lem:scalar-sol-nplus}).
Also, our vector network coding constructions from the previous sections are based on rank-metric codes,
but these codes can be seen as so-called \emph{lifted rank-metric codes}, which are special
subspace codes~\cite{silva_rank_metric_approach}.

In this section and the following ones, we explain
the simple formulation of these constructions, demonstrate how our
rank-metric code based constructions can be improved by using larger subspace codes,
and present a general question on subspace codes.
Finally, we show a multicast network with three messages in which a vector solution
outperforms the optimal scalar linear network solution, where the key is to use special
classes of subspace codes.

The formulation of the solutions with subspace codes is as follows.
Each message in the vector solution is a vector of length $t$.
The global coding vectors for a given edge consists of $h$ coding coefficients
$\Mat{A}_1,\Mat{A}_2 , \ldots, \Mat{A}_h$, which are~$t \times t$ matrices over $\F_q$.
The packet associated with
this edge is $( \Mat{A}_1 ~\Mat{A}_2 ~ \cdots ~ \Mat{A}_h ) \cdot (\vec{x}_1, \vec{x}_2, \dots ,\vec{x}_h)^T $. Instead
of choosing these $h$ coding coefficients based on lifted rank-metric codes, as in our
previous constructions, we can use a $t \times (ht)$ matrix over~$\F_q$
which is a basis for a $t'$-dimensional subspace of $\F_q^{ht}$, for some $t' \leq t$.
A receiver $R$ can recover the $h$ messages if the~$t \times (ht)$ matrices on $In(R)$ span
the $(ht)$-dimensional space~$\F_q^{ht}$. Each of these $t \times (ht)$ matrices is divided
into $h$ matrices from $\F_q^{t \times t}$ which are the global coding vectors on the edges.
As before, we have to distinguish between the local coding vectors
and the global coding vectors.
We omit the discussion on the local coding vectors since in our networks,
the local and global coding vectors are equal.

We will now consider some of the networks used
in our discussion and analyze them with respect to a solution with subspace codes.
We thereby consider only \emph{Grassmannian codes}, i.e., all subspaces have the same dimension.
Note that in some cases, better codes might be obtained
by using subspaces of different dimensions.
However, we will not consider them in this paper, as the
general idea can be understood from the Grassmannian codes, and the order of
magnitude does not change (which is the same as the one by the constructions
based on rank-metric codes). We should note that
if we append the $h$ coding coefficients which are $t \times t$ matrices in
the general framework of vector network coding, we will obtain $t \times (ht)$
matrices which are a basis of a subspace, and we will obtain a construction
based on subspace codes. However, to figure out which~$t \times t$ matrices
to take for this purpose is an almost impossible mission without going through the
related subspace codes.

Now, we consider the $(\epsilon,\ell)$-$\mathcal{N}_{2\ell,r,s}$ network. In such a network,
the sets of $\ell$ parallel edges carry
subspaces from $\Grassm{2\ell t,\ell t}$, while the $\epsilon$ parallel direct links from the unique source
to the receivers carry subspaces from $\Grassm{2\ell t,\epsilon t}$. The $(\ell t)$-dimensional
subspaces on the edges from the source to the middle layer form a code $\CC$ in $\Grassm{2\ell t,\ell t}$ such that
any two codewords (subspaces) of $\CC$ span a subspace of $\F_q^{2\ell t}$ whose dimension is
at least $(2\ell -\epsilon )t$, i.e., the minimum subspace distance of $\CC$ is at
least $(2\ell -2\epsilon )t$.

We now consider the solution for the $(\ell -1,\ell)$-$\mathcal{N}_{2\ell,r,3\ell-1}$ network.
The scalar solution and the vector solution are very similar for this network.
Recall that the optimal scalar solution (given in Lemma~\ref{lem:scalar-sol-nplus}) is obtained when we consider
that a node in the middle layer is transmitting $\ell$ blocks, each one with $2 \ell$ symbols from the alphabet $q_s$.
In the optimal scalar linear solution, each of these $\ell$ blocks forms an $\ell$-dimensional subspace of $\F_{q_s}^{2\ell}$ such
that any two such $\ell$-dimensional subspaces intersect in an at most $(\ell-1)$-dimensional subspace.
Hence, the subspace distance between two such subspaces is at least two, i.e., the largest such
set of subspaces consists of all the $\ell$-dimensional subspaces
of $\Grassms {2 \ell , \ell}$. The size of $\Grassms {2 \ell , \ell}$  is $\quadbinoms {2 \ell}{\ell}< 4q_s^{\ell^2}$ (see Section~\ref{subsec:prilim-subspaces}).
For the vector solution, we
use a subspace code $\CC$ in $\Grassm{2 \ell t , \ell t}$ with subspace distance
at least $2t$. The maximum size of such a code is larger than $q^{\ell (\ell -1) t^2 +\ell t}$ (see Theorem~\ref{thm:bound_sub_codes}).
Thus, we have that $q_s = {q^{(\ell-1)t^2/\ell + o(t)}}$, while $q_v=q^t$.

Constructions of large codes for this purpose can be found
for example in~\cite{Etzion2009ErrorCorrecting}. However, as said the improvement is not large since
asymptotically the subspace code obtained from an MRD code which was used for example in
Construction~\ref{constr:mrd-notfull-rank-comb-2} is optimal and can be improved by
at most a factor of two (see Theorem~\ref{thm:bound_sub_codes}).

\section{Analyzing Generalized Combination Networks with Vector Solutions}
\label{sec:analyzing}
In this section, we will compare the optimal scalar linear solution to our
vector solution
with respect to the alphabet size for some subfamilies of the generalized combination networks.

Consider first the unmodified $\mathcal{N}_{h,r,s}$ combination network.
It was proven in~\cite{RiisAhlswede-ProblemsNetworkCodingECC}
that the network has a scalar solution (not necessarily linear) if and only if a code over $\F_{q_s}$ with $q_s^h$ codewords and
minimum Hamming distance $r-s+1$ exists. A scalar linear solution exists for related
linear codes.
For a vector solution based on subspace codes, we need a code~$\mathbb{C}$ which is a subset
of $\Grassm{ht,t}$ such that any $s$ codewords of $\mathbb{C}$ span $\F_q^{ht}$. Such a vector solution
can be constructed from the scalar linear solution by using the companion matrix and its powers; and
it is equivalent to the scalar solution. In order to outperform the
scalar linear solution, we would need a larger such code.
However, no set of parameters for which such a code exists is known.

The next considered subfamily of generalized combination networks is the $(\epsilon,1)$-$\mathcal{N}_{h,r,s}$
network (Section~\ref{subsec:eps-1-network}). Three messages and $s=h$ will be considered
in the next section, where we will discuss also the other networks
in this subfamily and also related networks.

From the given framework of the vector network solution using subspace codes,
it is easy to verify that if we want to use error-correcting constant dimension codes
for our vector solution of the $(\epsilon,\ell)$-$\mathcal{N}_{h,r,s}$ network, with
$\ell \geq 1$, then it is required that $\frac{s-\epsilon}{\ell}$ equals two.
Hence, the number of receivers is $\binom{r}{2}$, as each receiver gets
its information from two distinct nodes from the middle layer;
otherwise the requirement is that at least three codewords will participate when
we examine the dimension spanned by the appropriate number of subspaces.
The two examples of networks for which our vector solution outperforms the optimal scalar
linear solution (see Section~\ref{sec:comb-extra-link-removed-receivers}) have this property, and $\frac{s-\epsilon}{\ell} = 2$.
Clearly, we must also have that ${\epsilon + \ell < h}$, since otherwise the information on the $\epsilon$ direct
links can always complete the information on the $\ell$ parallel links to the
ambient $h$-space. Another requirement is that ${h - 2 \ell \leq \epsilon}$, since otherwise the
cut at the receivers will be ${2 \ell + \epsilon <h}$, and hence the receiver will not be able to
recover $h$ messages, as $h$ is larger than the min-cut. Therefore, we have that
$h -2 \ell \leq \epsilon < h - \ell$, or equivalently $\epsilon < h - \ell \leq \epsilon + \ell$.
Since the $\epsilon$ direct links can always be chosen in a way that they complete the
information with the other $2 \ell$ links to the ambient $h$-space, it follows that
the value of $r$ depends only on the choice of the information on the $2 \ell$
links entering any given receiver. Each receiver should obtain a $(2\ell -\epsilon)$-subspace of $\F_{q_s}^h$.
Hence, for the scalar linear solution we require to have a code $\CC$ in the
Grassmannian $\cG_{q_s}  (h , \ell)$ such that its minimum subspace distance $d=2\delta$
satisfies $h -\epsilon  =  \ell + \delta$, i.e., $\delta = h - \ell - \epsilon$.
For the optimal scalar linear solution, we distinguish between two cases: $h- \ell \leq \ell$ and $h-\ell \geq \ell$.
If $h-\ell \geq \ell$, then our code $\CC$
has at least size ${ q_s^{(h-\ell) (2\ell -h+\epsilon +1) } }$  by Theorem~\ref{thm:bound_sub_codes}.
If $h-\ell \leq \ell$, then our code $\CC$
has at least size ${ q_s^{\ell (\epsilon +1) } }$  by Theorem~\ref{thm:bound_sub_codes}.

For our vector solution, our best code $\CC$ is in the Grassmannian $\Grassm {h t , \ell t}$,
and with the same reasons, its minimum subspace distance is $2(h -\ell - \epsilon)t$.
We distinguish again between the same two cases: $h- \ell \leq \ell$ and $h-\ell \geq \ell$.
If $h-\ell \geq \ell$, then our code $\CC$
has size at least $q^{(h-\ell)t (2\ell t-ht+\epsilon t +1) }$  by Theorem~\ref{thm:bound_sub_codes}.
If $h-\ell \leq \ell$, then our code $\CC$
has size at least $ q^{\ell t (\epsilon t+1) }$  by Theorem~\ref{thm:bound_sub_codes}.
The following theorem summarizes the minimal order of $q_s$.
\begin{theorem} %[Order of Scalar Solutions for Generalized Network]
\label{thm:scalar-sol-gen-network}
For the $(\epsilon,\ell)$-$\mathcal{N}_{h,r,2 \ell + \epsilon}$ network, there exists a vector solution of dimension $t$ over a field of size $q$, i.e., $q_v=q^t$, and
the scalar linear solution has field size at least $q_s$ (for the following given $r$),
\begin{enumerate}
\item if $h-\ell \leq \ell$ and $\epsilon=0$, then $r = q^{\ell t}$ and ${q_s=q^t}$.

\item if $h-\ell \leq \ell$ and $\epsilon \neq 0$, then $r = q^{(\ell t)(\epsilon t+1)}$
and $q_s = q^{\epsilon t^2 /(\epsilon +1) + o(t)}$.

\item if $h-\ell \geq \ell$ and $2\ell -h+\epsilon=0$, then $r = q^{(h-\ell )t}$ and $q_s = q^t$.

\item if $h-\ell \geq \ell$ and $2\ell -h+\epsilon \neq 0$, then $r = q^{(h - \ell )t(2 \ell t -ht+ \epsilon t +1)}$
and $q_s = q^{(2\ell -h +\epsilon)t^2/(2\ell-h+\epsilon+1) + o(t)}$.
\end{enumerate}
\end{theorem}
%\awcomment{Is it really $r\geq ...$ and not $r \leq ..$?}

To obtain the largest possible gap between $q_v$ and $q_s$ using our methods, we have to optimize the order in the cases
given in Theorem~\ref{thm:scalar-sol-gen-network}, which do not depend only on $t$, i.e., case~2 and case~4.

\begin{enumerate}
\item[2)] If $h-\ell \leq \ell$ and $\epsilon \neq 0$, then we have to maximize $q^{\epsilon t^2 /(\epsilon +1)}$,
i.e., for a given $h$, we must have the largest possible $\epsilon$. Since $\ell + \epsilon < h$,
it follows that we should have $\epsilon = h - \ell -1$, and $\ell$ should be small as possible.
\begin{enumerate}
\item If $h$ is even, then the smallest $\ell$ is obtained when $h =2 \ell$ (since $h \leq 2\ell$),
i.e., $\ell = \frac{h}{2}$. Hence, $\epsilon$ gets the maximum value when ${\epsilon =\ell -1 = \frac{h}{2}-1}$.
Therefore, the largest gap occurs when $q_s = q^{(h-2)t^2/h + o(t)}$, and the network that
achieves the largest gap is the
$(\ell-1,\ell)$-$\mathcal{N}_{2\ell,r,3\ell -1}$ network.

\item If $h$ is odd, then the smallest $\ell$ is obtained when $h =2 \ell-1$ (since $h \leq 2\ell$),
i.e., $\ell = \frac{h+1}{2}$.
Thus, $\epsilon$ has to be maximized, and
the largest gap occurs when $q_s = q^{(h-3)t^2/(h-1) + o(t)}$.
The network that achieves this largest gap is the
$(\ell-2,\ell)$-$\mathcal{N}_{2\ell-1,r,3\ell-2}$ network.
\end{enumerate}

\item[4)] If $h-\ell \geq \ell$ and $2 \ell -h +\epsilon \neq 0$,
then we have to maximize $q^{(2\ell -h +\epsilon)t^2/(2\ell-h+\epsilon+1)}$,
i.e., for a given $h$, we must have the largest possible value for $2\ell -h +\epsilon$. Since $\ell + \epsilon < h$,
it follows that we should have $\epsilon = h - \ell -1$, and hence we have to maximize
$2\ell -h +\epsilon=2\ell -h +h - \ell -1=\ell-1$.
\begin{enumerate}
\item If $h$ is even, then the largest $\ell$ is obtained for $h =2 \ell$ (since $h \geq 2\ell$),
i.e., $\ell = \frac{h}{2}$. Therefore, $2\ell -h +\epsilon$ is maximized
for ${2\ell -h +\epsilon=2\frac{h}{2}-h+h-\frac{h}{2}-1=\frac{h}{2}-1}$.~Thus,
the largest gap occurs when $q_s=q^{(h-2)t^2/h + o(t)}$, and the network that
achieve the largest gap is the
$(\ell-1,\ell)$-$\mathcal{N}_{2\ell,r,3\ell -1}$ network (the same one as before,
since the intersection of the two cases is when $h -\ell = \ell$).

\item If $h$ is odd, then the largest gap is obtained for $h =2 \ell+1$ (since $h \geq 2\ell$),
i.e., $\ell = \frac{h-1}{2}$. Hence, $2\ell -h +\epsilon$ is maximized
for $2\ell -h +\epsilon=2\frac{h-1}{2}-h+h-\frac{h-1}{2}-1=\frac{h-1}{2}-1$.
It implies that the largest gap occurs when $q_s=q^{(h-3)t^2/(h-1) + o(t)}$, and the network that
achieves the largest gap is the
$(\ell-1, \ell)$-$\mathcal{N}_{2\ell+1,r,3 \ell-1}$ network.
\end{enumerate}
\end{enumerate}

\section{Scalar and Vector Solutions for General Networks}
\label{sec:network-3-messages}

In the previous sections, we have discussed specific subfamilies of the generalized
combination networks. The main goal was to find networks that, to our knowledge,
achieve the largest gap in the alphabet size between a vector solution
and the optimal scalar linear solution. In this section, we discuss the requirements
needed for a scalar linear solution and the requirements needed for a vector solution
for another subfamily of the generalized combination networks. As before,
we aim to find the smallest alphabet size for
which such a solution exists. Indeed, we will start with the most general network,
the $(\epsilon,\ell)$-$\mathcal{N}_{h,r,\alpha \ell + \epsilon}$ network, to
provide the requirements for its solution. However, the code required for
such a solution cannot be constructed generally for a large set of parameters.
Hence, also in this section, we will continue in a few steps
from simpler subfamilies to more evolved ones.
The reason is that a network coding problem for
simple subfamilies might be feasible to solve, while it might be too difficult
to provide the solution for the the most general
networks. One of the highlights of this section is a network
with three messages in which vector network coding outperforms scalar network coding.
This is one of the parameters we could not obtain in our previous constructions. For two messages,
unfortunately, this problem remains open. The vector solution for three messages
raises an interesting open question on subspace codes.

The most general network is the $(\epsilon,\ell)$-$\mathcal{N}_{h,r,\alpha \ell + \epsilon}$ network,
where $\epsilon \geq 0$ and $\ell \geq 1$. To avoid trivial solutions (i.e., routing), we
must have that $\ell+\epsilon < h$, and to avoid unsolvable networks
by the min-cut/max-flow condition, we must have that ${\alpha \ell + \epsilon \geq h}$.
This is summarized in the following theorem.
\begin{theorem}
\label{thm:trivialORno}
The $(\epsilon,\ell)$-$\mathcal{N}_{h,r,\alpha \ell + \epsilon}$ network
has a trivial solution if $\ell + \epsilon \geq h$, and it has
no solution if $\alpha \ell + \epsilon < h$.
\end{theorem}
We start with the scalar linear solution for the general network. Each set of parallel $\ell$
links carries $\ell$ vectors which span a subspace of dimension at most $\ell$ from the $h$-space $\F_{q_s}^h$.
The subspaces on these $\ell$ parallel links define a subspace code in~$\cP_{q_s}(h)$
(the set of all subspaces from $\F_{q_s}^h$) such that
any subset of $\alpha$ codewords spans a subspace of $\F_{q_s}^h$ whose dimension
is at least $h-\epsilon$. We continue with the vector solution for the general network. Each set of parallel $\ell$
links carries a vector which spans a subspace of dimension at most $\ell t$ from the $(ht)$-dimensional subspace $\F_q^{ht}$.
The subspaces on these $\ell$ parallel links define a subspace code in $\cP_q(ht)$ such that
any subset of $\alpha$ codewords (subspaces) spans a subspace of $\F_q^{ht}$ whose dimension
is at least $ht-\epsilon t$.
The requirements on codes or subspace codes for the scalar linear solution
and the subspace codes for the vector solution are too general.
We suggest to split the problem into a few questions
related to the different subfamilies of
the generalized combination network.

\subsection{Scalar solution for the $(\epsilon,1)$-$\mathcal{N}_{h,r,\alpha + \epsilon}$ network}

We start with the $(\epsilon,1)$-$\mathcal{N}_{h,r,\alpha + \epsilon}$ network.
This network is very similar to the $\mathcal{N}_{h,r,\alpha}$ combination network.
The only difference between these two classes of networks are the $\epsilon$ direct links
from the source to each receiver.
For the optimal scalar linear solution, we need an $h \times r$ matrix over $\F_q$ such that
any $\alpha$ columns span a subspace of~$\F_{q_s}^h$ whose dimension is at least $h - \epsilon$.
This is usually not a simple problem. For $\epsilon =0$ this is equivalent to the error-correcting
code problem as was defined in~\cite{RiisAhlswede-ProblemsNetworkCodingECC}, i.e.,
constructing a code of length $r$ with $q_s^h$ codewords
with minimum Hamming distance $r-\alpha +1$ for a field of smallest possible size $q_s$. This can be generalized for
any $\epsilon$ such that $0 \leq \epsilon \leq h-2$.
A scalar solution (linear or nonlinear) over $\F_{q_s}$ for the $(\epsilon,1)$-$\mathcal{N}_{h,r,\alpha + \epsilon}$ network
exists if and only if there exists a code over $\F_{q_s}$ of length $r$ with $q_s^h$ codewords
and minimum Hamming distance $r-\alpha-\epsilon+1$. Hence, the addition of the
direct edges from the source to the receivers enables us to use longer codes with smaller
alphabet size for the scalar solution, but the type of the coding problem remains the same.
Therefore, the question on a suitable construction is a standard problem in error-correcting codes.

\subsection{Vector solution for the $(\epsilon,1)$-$\mathcal{N}_{h,r,\alpha + \epsilon}$ network}

For the vector solution, we need something different, stated in the following question.
\begin{question}
\label{ques:vec-sol-subspace}
What is the largest set of $r$ subspaces of dimension at most $t$ of the space $\F_q^{ht}$ such that
any subset of $\alpha$ subspaces spans a subspace of dimension at least $(h-\epsilon)t$?
\end{question}

While the scalar linear solution poses a classic coding problem for this network,
the vector solution poses a completely new coding problem on subspace codes.
In fact, while we have lot of knowledge on the solution for the coding problem
posed by the scalar linear solution, we hardly know anything about an answer
to Question~\ref{ques:vec-sol-subspace}.
As described in previous sections, if $\alpha =2$, then Question~\ref{ques:vec-sol-subspace}
reduces to the problem of designing error-correcting subspace codes.
We discuss now a more specific subfamily of these networks.

The $(1,1)$-$\mathcal{N}_{h,r,h + 1}$ network, with $h \geq 3$, is the
simplest known modification of the combination networks from all our new networks
which exhibit the advantage of vector network coding on scalar network coding
with respect to the alphabet size. The $(1,1)$-$\mathcal{N}_{h,r,h + 1}$ network
consists of the unmodified combination network $\mathcal{N}_{h,r,h}$ with an additional direct link from the source
to each receiver.

\subsection{Scalar solution for the $(1,1)$-$\mathcal{N}_{3,r,4}$ network}

In this subsection, we discuss the case $h=3$, i.e., the source has three messages
(see also Fig.~\ref{fig:comb-net-extra}).
Given a receiver~$R$, it obtains its information from three nodes in the middle
layer and from the direct link from the source. For the scalar linear solution,
each such link has a coding vector of length three.
The receiver is able to recover the three messages, if its four received vectors
span a subspace of dimension three. This implies that the three links from the middle
layer carry three vectors which span a subspace of $\F_{q_s}^{4}$ whose dimension
is at least two. To ensure that each receiver
will receive three vectors which span such a subspace, we have to guarantee that
on the links between the source and the nodes of the middle layer, no three links will contain a vector
which is contained in the same one-dimensional subspace (two such links can have such
vectors). Thus, for a scalar linear solution, we must have that
$r \leq 2\quadbinoms{3}{1} = 2\frac{q_s^3-1}{q_s-1} = 2(q_s^2+q_s+1)$.
Fortunately, a related assignment of the coefficients for the coding vectors of
the edges induces a scalar solution with $r=2(q_s^2+q_s+1)$.

\subsection{Vector solution for the $(1,1)$-$\mathcal{N}_{3,r,4}$ network}

We continue with a vector solution for this network.
Assume now that the three messages are vectors of length~$t$ over~$\F_q$. Now,
the coefficients on each link are three $t \times t$ matrices over~$\F_q$.
Each such three coefficients on a link can be represented as
a $t \times (3t)$ matrix. The receivers can recover the three messages if
any three such matrices on the links between the source and the nodes
of the middle layer span a subspace of~$\F_q^{3t}$ whose dimension is at least $2t$.
Our problem is to find the largest possible $r$ with this property, and it can
be formulated by the following coding problem in the Grassmannian $\Grassm{3t,t}$:
Find the largest set of subspaces from~$\Grassm{3t,t}$ such that any three
subspaces of the set span a subspace of dimension at least $2t$.
If the size of this set $r$ is greater than $2(q_s^2+q_s+1)$, then our vector solution
outperforms the optimal scalar linear solution in the $(1,1)$-$\mathcal{N}_{3,r,4}$ network.

We consider now a specific example which can be generalized.
Assume $q_s=4=2^2$, i.e., $r \leq 2 \cdot (4^2+4+1)=42$, and consider the vector solution,
where the messages are binary vectors of length $t=2$.
Hence, the edges will carry
two-dimensional subspaces of $\F_2^6$. Vector network coding will outperform optimal
scalar linear network coding if there exist more than 42 two-dimensional subspaces
of $\F_2^6$ such that any three two-dimensional subspaces will span at least a four-dimensional
subspace of $\F_2^6$, so they will be completed by the direct link from the source to the receiver.

Let $\beta$ be a primitive element in $\F_{2^4}$ satisfying $\beta^4=\beta +1$.
%With the 51 two-dimensional subspaces of $\F_2^6$, we form the following sets:
We form the following set of 51 two-dimensional subspaces of $\F_2^6$:
$\{ \langle 01 \beta^i , 10 \beta^{i+1} \rangle ~:~ 0 \leq i \leq 14 \}$,
$\{ \langle 01 \beta^i , 10 \beta^{i+2} \rangle ~:~ 0 \leq i \leq 14 \}$,
$\{ \langle \{ 01 \beta^i , 10 \beta^{i-1} \rangle ~:~ 0 \leq i \leq 14 \}$,
$\{ \langle 00 \beta^i , 00 \beta^{i+5} \rangle ~:~ 0 \leq i \leq 4   \}$, and $\{ \langle 100000,010000 \rangle \}$.
One can easily verify that any three of these 51 two-dimensional subspaces
span at least a four-dimensional subspace and hence, our vector solution
outperforms the optimal scalar linear solution in this case. Recently, 82
two-dimensional subspaces
of $\F_2^6$, such that any three two-dimensional subspaces will span at least a four-dimensional
subspace of $\F_2^6$, were found~\cite{EOO18}.

\section{Conclusions and problems for Future Research}
\label{sec:conclusion}

We have shown that our vector solutions outperform the optimal scalar linear solution
in the alphabet size for several generalizations of the combination networks.
The key is the use of subspace codes, particularly subspace codes derived from rank-metric codes.
Our networks and codes imply a gap of size $q^{(h-2)t^2/h + o(t)}$ for even $h \geq 4$,
and a gap of size $q^{(h-3)t^2/(h-1) + o(t)}$
for any odd $h \geq 5$, between the alphabet size {of our} vector solution
and {the optimal} scalar linear solution for multicast networks, where $h$ is the number of messages and $t$ is the length of the vectors.
Clearly, a vector solution can be translated to a scalar nonlinear
 solution. Therefore, our results also imply a gap of size $q^{(h-2)t^2/h + o(t)}$ for any even $h \geq 4$,
and a gap of size $q^{(h-3)t^2/(h-1) + o(t)}$
for any odd $h \geq 5$, between the field size in \emph{linear} and \emph{nonlinear} scalar network coding
for multicast networks. We have also proved
the existence of a multicast network with three messages in which the vector solution
outperforms the scalar linear solution.

There are many interesting questions which remain open after our discussion.
Some of these open questions for future research are briefly outlined and discussed as follows.
\begin{enumerate}
\item Can the $\mathcal{N}_{h,r,h}$ combination network have a vector solution
based on an $\MRDlinq{t\times t, t}$ code which outperforms the optimal scalar linear solution?
We conjecture that the answer is NO. Clearly,
if the MRD code consists of a companion matrix and its powers, then this is not
possible since each power of the companion matrix can be translated to the related
element in the finite field and the vector solution can be translated to a scalar linear solution.
Are all the $\MRDlinq{t\times t, t}$ codes equivalent in this sense?
We conjecture that the answer is YES.

\item Can the $\mathcal{N}_{h,r,h}$ combination network have a vector solution based
on subspace codes which outperforms the optimal scalar linear solution? If such a code $\mathbb{C}$ with $r$ codewords exists, then
any $h$ $t$-dimensional subspaces of~$\mathbb{C}$ have to span the ambient space $\F_q^{ht}$.
We have shown that the answer is negative for $h=2$ and we conjecture it is negative for all $h$.

\item Can any $\mathcal{N}_{h,r,s}$ combination network have a vector solution which
outperforms the optimal scalar linear solution? Since the scalar solution of the $\mathcal{N}_{h,r,s}$ combination network
uses an error-correcting code, this question is of a great interest.
The scalar solution for the $\mathcal{N}_{h,r,h}$ combination network uses an MDS code.
A vector solution which outperforms the scalar solution implies subspace codes which go beyond
the MDS bound. For some codes, we can prove that a vector solution cannot outperform
the optimal scalar linear solution.

\item Does any $(0,\ell)$-$\mathcal{N}_{h,r,s}$ network, $\ell >1$, have a vector solution which outperforms
the scalar solution? If $s=2 \ell$ this is related to some interesting questions
on subspace codes, where the most interesting case
is related to spreads and partial spreads in projective geometry.

\item
Does there exist a multicast network with \emph{two} messages in which a vector solution
outperforms the optimal scalar linear solution?

\item
For multicast networks with $h = 3$ messages which have a vector solution of
dimension $t$ over $\F_q$, what is the largest alphabet size required for
a solution with scalar linear network coding?
It was pointed out by
Ronny Roth~\cite{Roth16} that by a probabilistic argument,
there exists a set with at least $q^{t^2/3}$ $t$-dimensional
subspace of~$\F_q^{3t}$ for which any three subspaces span a subspace of dimension at least~$2t$.
This probabilistic argument can be generalized for any $h$.
Is this an optimal solution?
%Note that for $h=3$, the $(1,1)$-$\mathcal{N}_{3,r,4}$ network achieves
%a gap of $q^{t^2/3 + o(t)}$, but we do not have a construction of the network code.
It is a future task to find a construction of such a code and possibly a larger one.

\item For multicast networks which have a vector solution of dimension $t$ over $\F_q$, is there an algorithm which transforms the
vector solution into a scalar linear solution?
What is the alphabet size required
by this transformation? We believe that
the minimum alphabet size for scalar network coding is at most $q^{(h-2) t^2/h +o(t)}$ for even $h\geq 4$
and at most $q^{(h-3) t^2/(h-1) +o(t)}$ for odd $h\geq 5$,
but an algorithm for any larger alphabet might be interesting to begin with,
e.g., an alphabet of size $q^{t^2}$ seems to be an excellent achievement.

\item Is there a network with $h$ messages in which exactly $h$ edge disjoint paths
are used (for network coding) from the source
to each receiver, and on which vector network coding
outperforms scalar linear network coding?
Note that our constructions use more than $h$ paths. In other words, can
vector network coding outperform scalar linear network coding in a network
in which the deletion of any edge will destroy the multicast property?

\item What is the largest possible gap between the field size in scalar nonlinear network coding
and scalar linear network coding for multicast networks? This gap is at least the size of the gap between the
vector solution and the optimal scalar linear solution. However, is it possible
to obtain a larger gap between nonlinear and linear scalar network coding?
Dougherty \emph{et al.}~\cite{DoughertyFreilingZeger-LinearitySolvabilityMulticastNetworks_2004} proved that for any integer $h \geq 3$, there exists
a multicast network with $h$ messages that has a binary solution but does not have a
binary linear solution. This result implies that there is no vector solution with vectors
of length one over~$\F_2$. Can this result can be generalized to any non-binary alphabet?
Can it be also generalized to vectors of length greater than one? What is the {largest} possible gap
in the alphabet size between nonlinear scalar network coding and vector network coding?

\item For given $h \geq 3$, $\alpha \geq 3$, and $1 \leq \epsilon \leq h-2$, what is the size of the largest code in $\Grassm {ht,\ell t}$
such that any $\alpha$ codewords span a subspace of dimension at least $(h-\epsilon)t$?
This question has an immediate application for $h=3$, $\alpha =3$, and $\epsilon =1$ in constructing networks
with three messages for which vector network coding outperforms scalar coding.
Also, it is interesting to see the related
results in the performance of vector network coding for $h > 3$. This is a generalization of the
standard coding problem for subspace codes. For $\alpha=2$,
such a code is a subspace code with minimum subspace
distance ${2 (h -\ell - \epsilon )}$. This problem was considered recently in~\cite{EtZh18}.

\end{enumerate}

\section*{Acknowledgement}

The authors are indebted to the two anonymous reviewers whose constructive comments have
made an important contribution to the presentation of this paper.
Especially, one of the reviewers was generous enough to mark any spot in the text that needed
some alteration.

The authors would like to thank Ronny Roth for many valuable discussions and for constructing
the right subspace code for three messages using a probabilistic method.
The authors would also like to thank Netanel Raviv for pointing out an error in an earlier version.

% Generated by IEEEtranS.bst, version: 1.13 (2008/09/30)

\newpage

\begin{IEEEbiographynophoto}{Tuvi Etzion} (M'89--SM'94--F'04) was born in Tel Aviv, Israel,
in 1956. He received the B.A., M.Sc., and D.Sc. degrees from the
Technion - Israel Institute of Technology, Haifa, Israel, in 1980,
1982, and 1984, respectively.

From 1984 he held a position in the Department of Computer Science
at the Technion, where he now holds the Bernard Elkin Chair
in Computer Science. During the
years 1985-1987 he was Visiting Research Professor with the
Department of Electrical Engineering - Systems at the University
of Southern California, Los Angeles. During the summers of 1990
and 1991 he was visiting Bellcore in Morristown, New Jersey.
During the years 1994-1996 he was a Visiting Research Fellow in
the Computer Science Department at Royal Holloway University of London, Egham,
England. He also had several visits to the Coordinated Science
Laboratory at University of Illinois in Urbana-Champaign during
the years 1995-1998, two visits to HP Bristol during the summers
of 1996, 2000,  a few visits to the Department of Electrical
Engineering, University of California at San Diego during the
years 2000-2017, and several visits to the Mathematics Department
at Royal Holloway University of London, during the years
2007-2017.

His research interests include applications of discrete
mathematics to problems in computer science and information
theory, coding theory, network coding, and combinatorial designs.

Dr. Etzion was an Associate Editor for Coding Theory for the IEEE
Transactions on Information Theory from 2006 till 2009.
From 2004 to 2009, he was an Editor for the
Journal of Combinatorial Designs.
From 2011 he is an Editor for Designs, Codes, and Cryptography.
and from 2013 an Editor for Advances of Mathematics in Communications.
\end{IEEEbiographynophoto}

\begin{IEEEbiographynophoto}{Antonia Wachter-Zeh}
	(S’10--M'14) is an Assistant Professor at the Technical University of Munich (TUM), Munich Germany.
	She received an equivalent of the B.S. degree in electrical engineering in 2007 from the University of Applied Science Ravensburg, Germany, and the M.S. degree
	in communications technology in 2009 from Ulm University, Germany.
	She obtained her Ph.D. degree in 2013 at the Institute of Communications Engineering, University of Ulm, Germany and at the
	Institut de recherche mathématique de Rennes (IRMAR), Université de Rennes 1, Rennes, France.
	From 2013 to 2016, she was a postdoctoral researcher at the Technion---Israel Institute of Technology, Haifa, Israel.
	Her research interests are coding and information theory and their application to storage, communications, and security.
\end{IEEEbiographynophoto}

\end{document}

%% file: my_defs-v1.tex
\newtheorem{theorem}{Theorem}

\newtheorem{lemma}{Lemma}
\newtheorem{corollary}{Corollary}

\newtheorem{remark}{Remark}

\newtheorem{construction}{Construction}
\newtheorem{question}{Question}

 \newcommand{\qed}{\hfill \mbox{\raggedright \rule{.07in}{.1in}}}

\newcommand{\Fqm}{\ensuremath{\mathbb F_{q^m}}}

\newcommand{\Fq}{\ensuremath{\mathbb F_{q}}}

\newcommand{\F}{\ensuremath{\mathbb F}}
\newcommand{\CC}{\ensuremath{\mathbb C}}

\newcommand{\myset}[1]{\mathcal{#1}}

\newcommand{\OCompl}[1]{\ensuremath{\mathcal{O}({#1})}}

\SetAlgoCaptionSeparator{.} % Such that it's Algorithm(). not :
\DeclareMathOperator{\rk}{rk}

\renewcommand{\vec}[1]{\ensuremath{\mathbf{#1}}}
\newcommand{\Mat}[1]{\ensuremath{\mathbf{#1}}}

\newcommand{\x}{\mathbf x}
\newcommand{\y}{\mathbf y}

\newcommand{\A}{\Mat{A}}
\newcommand{\B}{\Mat{B}}
\newcommand{\C}{\Mat{C}}

\newcommand{\I}{\mathbf I}

\newcommand{\0}{\vec{0}}

\newcommand{\mycode}[1]{\ensuremath{\mathcal{#1}}}

\newcommand{\codelinearArb}[1]{\ensuremath{[#1]}}

\newcommand{\codelinearRank}[1]{\ensuremath{[#1]_q^\fontmetric{R}}}

\newcommand{\MRDlinq}[1]{\ensuremath{\mycode{MRD}[#1]_q}}

\newcommand{\fontmetric}[1]{\mathsf{#1}}

\newcommand{\Subspacedist}[1]{d_s(#1)}

\newcommand{\myspace}[1]{\mathcal{#1}}

\newcommand{\Grassm}[1]{\myspace{G}_q(#1)}
\newcommand{\Grassms}[1]{\myspace{G}_{q_s}(#1)}

\newcommand{\cP}{\mathcal{P}}
\newcommand{\cG}{\mathcal{G}}

\newcommand{\quadbinom}[2]{\ensuremath{
{#1
\brack
#2}_q
}}

\newcommand{\quadbinoms}[2]{\ensuremath{
{#1
\brack
#2}_{q_s}
}}

%% file: comb-network.tex
\centering
\tikzsetnextfilename{2-butterflies}
\begin{tikzpicture}[scale = 1]
	% one butterfly
	\node[mycircle,label=right:{$\vec{x}_1,\dots,\vec{x}_h$}] (sourcex) {} ; 
	\node[mycircle,below left=30pt and 40pt of sourcex] (middle0) {} ;
	\node[mycircle,below left=30pt and 20pt of sourcex] (middle1) {} ;
	\node[mycircle,below left=30pt and 0pt of sourcex] (middle2) {} ;
	\node[rectangle,right = 10pt of middle2](text1){$\dots$};
	\node[mycircle,below left=30pt and -60pt of sourcex] (middle3) {} ;
	\node[mycircle,below left=30pt and -80pt of sourcex] (middle4) {} ;
	\node[rectangle,right = 5pt of middle4](text2){$r$ nodes};

	\draw[black,->,very thick] (sourcex.south) -- (middle0.north);
	\draw[black,->,very thick] (sourcex.south) -- (middle1.north);
	\draw[black,->,very thick] (sourcex.south) -- (middle2.north);
	\draw[black,->,very thick] (sourcex.south) -- (middle3.north);
	\draw[black,->,very thick] (sourcex.south) -- (middle4.north);

	\draw [-, very thick,black] (2.2,-2.3) arc (160:50:10pt);
	\draw [-, very thick,black] (-1.8,-2.4) arc (140:30:13pt);
	\draw [-, very thick,black] (-2.5,-2.3) arc (150:20:10pt);
	
	\node[mycircle,below left=30pt and 10pt of middle0] (rec1) {} ;
	\node[mycircle,below left=30pt and -20pt of middle0] (rec2) {} ;
	\node[mycircle,below left=30pt and -150pt of middle0] (reclast) {} ; %,label=below:{R$\binom{r}{s}$}] (reclast) {} ;
	
	\draw[black,->,very thick] (middle0.south) -- (rec1.north);
	\draw[black,->,very thick] (middle1.south) -- (rec1.north);
	\draw[black,->,very thick] (middle0.south) -- (rec2.north);
	\draw[black,->,very thick] (middle2.south) -- (rec2.north);
	\draw[black,->,very thick] (middle3.south) -- (reclast.north);
	\draw[black,->,very thick] (middle4.south) -- (reclast.north);
	\node[rectangle,below right = 7pt and 7pt of middle4](text2){$s$ edges};
	\node[rectangle,below right = 28pt and 20pt of middle4](text2){$\binom{r}{s}$ receivers};
	
	\node[rectangle,right = 23pt of middle2](blub){$\dots$};
		\draw[black,->,very thick] (blub.south) -- (reclast.north);
	
\end{tikzpicture}

%% file: network-ell-tau.tex
\centering
\tikzsetnextfilename{2-butterflies}

%\tikzset{EdgeStyle/.append style = {->, bend left} }
  
\begin{tikzpicture}[scale = .9]
	% one butterfly
	\node[mycircle,label=right:{$\vec{x}_1,\dots,\vec{x}_h$}] (sourcex) {} 
	edge[bend left=5, very thick,->] (middle0.north)
	edge[bend right=5, very thick,->] (middle0.north)
	edge[bend left=5, very thick,->] (middle1.north)
	edge[bend right=5, very thick,->] (middle1.north)
	edge[bend left=5, very thick,->] (middle2.north)
	edge[bend right=5, very thick,->] (middle2.north)
	edge[bend left=5, very thick,->] (middle3.north)
	edge[bend right=5, very thick,->] (middle3.north)
	edge[bend left=5, very thick,->] (middle4.north)
	edge[bend right=5, very thick,->] (middle4.north)
	
	edge[bend right=40, very thick,->] (rec1.west)
	edge[bend right=35, very thick,->] (rec1.west)
	edge[bend left=40, very thick,->] (rec2.east)
	edge[bend left=35, very thick,->] (rec2.east)
	edge[bend right=20, very thick,->] (reclast.west)
	edge[bend right=15, very thick,->] (reclast.west)
	; 
	
	\draw [-, very thick,black] (.25,-1) arc (-110:-40:10pt);
	\draw [-, very thick,black] (-.03,-1) arc (-100:-40:10pt);
	\draw [-, very thick,black] (-2.4,-.9) arc (-160:-60:10pt);
	
	\draw [-, very thick,black] (1.7,-1) arc (-110:-40:10pt);
	\draw [-, very thick,black] (0.9,-.85) arc (-110:-40:10pt);
	\draw [-, very thick,black] (-1.45,-.7) arc (-160:-100:10pt);
	\draw [-, very thick,black] (-0.9,-.7) arc (-140:-70:10pt);
	\draw [-, very thick,black] (-0.45,-.7) arc (-120:-50:10pt);
	
	\draw [-, very thick,black] (1.93,-2) arc (-110:-40:10pt);
	\draw [-, very thick,black] (2.4,-2) arc (-110:-40:10pt);
	\draw [-, very thick,black] (-1.2,-2) arc (-140:-60:10pt);
	\draw [-, very thick,black] (-2.2,-2.1) arc (-160:-90:10pt);
	\draw [-, very thick,black] (-1.8,-2.2) arc (-110:-60:10pt);
	\draw [-, very thick,black] (-2.5,-2) arc (-160:-90:10pt);

	\draw [-, very thick,black] (2.1,-2.8) arc (170:60:20pt);
	
	\node[rectangle, below right=6.5pt and 47pt of sourcex](textell){\textcolor{black}{${\ell}$}};
	
	\node[rectangle, below left=6.5pt and 48pt of sourcex](textell2){\textcolor{black}{${\epsilon}$}};
	
	\node[mycircle,below left=30pt and 40pt of sourcex] (middle0) {} 
	edge[bend left=5, very thick,->] (rec1.north)
	edge[bend right=5, very thick,->] (rec1.north)
	edge[bend left=5, very thick,->] (rec2.north)
	edge[bend right=5, very thick,->] (rec2.north)
	
	; 
	\node[mycircle,below left=30pt and 20pt of sourcex] (middle1) {} 
	edge[bend left=5, very thick,->] (rec1.north)
	edge[bend right=5, very thick,->] (rec1.north);
	\node[mycircle,below left=30pt and 0pt of sourcex] (middle2) {} 
	edge[bend left=5, very thick,->] (rec2.north)
	edge[bend right=5, very thick,->] (rec2.north);
	\node[rectangle,right = 10pt of middle2](text1){$\dots$}
	;
	\node[mycircle,below left=30pt and -60pt of sourcex] (middle3) {} 
	edge[bend left=5, very thick,->] (reclast.north)
	edge[bend right=5, very thick,->] (reclast.north);
	\node[mycircle,below left=30pt and -80pt of sourcex] (middle4) {}
	edge[bend left=5, very thick,->] (reclast.north)
	edge[bend right=5, very thick,->] (reclast.north) ;
	\node[rectangle,right = 5pt of middle4](text2){$r$ \small{middle layer nodes}};

		\node[rectangle,below right = 2pt and 15pt of middle4](text2){$s=2\ell + \epsilon$}; 
		\node[rectangle,below right = 12pt and 20pt of middle4](text2){\small{incoming edges}};
		\node[rectangle,below right = 30pt and 27pt of middle4](text2){$\binom{r}{\alpha}$ \small{receivers}};
	
%	\draw[black,->,very thick] (sourcex.south) -- (middle0.north);
%	\draw[black,->,very thick] (sourcex.south) -- (middle1.north);
%	\draw[black,->,very thick] (sourcex.south) -- (middle2.north);
%	\draw[black,->,very thick] (sourcex.south) -- (middle3.north);
%	\draw[black,->,very thick] (sourcex.south) -- (middle4.north);
	
	\node[mycircle,below left=30pt and 10pt of middle0] (rec1) {} ;
	\node[mycircle,below left=30pt and -20pt of middle0] (rec2) {} ;
	\node[mycircle,below left=30pt and -150pt of middle0] (reclast) {} ; %,label=below:{R$\binom{r}{s}$}] (reclast) {} ;
	
%	\draw[black,->,very thick] (middle0.south) -- (rec1.north);
%	\draw[black,->,very thick] (middle1.south) -- (rec1.north);
%	\draw[black,->,very thick] (middle0.south) -- (rec2.north);
%	\draw[black,->,very thick] (middle2.south) -- (rec2.north);
%	\draw[black,->,very thick] (middle3.south) -- (reclast.north);
%	\draw[black,->,very thick] (middle4.south) -- (reclast.north);
%	\node[rectangle,below right = 7pt and 7pt of middle4](text2){$s$ edges};
	
\end{tikzpicture}

%% file: comb-network-extra-link.tex
\centering
\tikzsetnextfilename{2-butterflies}
\begin{tikzpicture}[scale = 1]
	% one butterfly
	\node[mycircle,label=right:{$\vec{x}_1,\vec{x}_2,\vec{x}_3$}] (sourcex) {}

		edge[bend right=5, very thick,->] (middle0.north)
		edge[bend right=5, very thick,->] (middle1.north)
		edge[bend right=5, very thick,->] (middle2.north)
		edge[bend left=5, very thick,->] (middle3.north)
%		edge[bend right=5, very thick,->] (middle3.north)
		edge[bend left=5, very thick,->] (middle4.north)
%		edge[bend right=5, very thick,->] (middle4.north)
		
		edge[bend right=50, very thick,->] (rec1.west)
		edge[bend left=30, very thick,->] (rec2.east)
		edge[bend right=25, very thick,->] (reclast.west)
		;

	\node[mycircle,below left=30pt and 40pt of sourcex] (middle0) {} 
	edge[bend left=5, very thick,->] (rec1.north)
%	edge[bend right=5, very thick,->] (rec1.north)
%	edge[bend left=5, very thick,->] (rec2.north)
	edge[bend right=5, very thick,->] (rec2.north)
	
	;

	\node[mycircle,below left=30pt and 20pt of sourcex] (middle1) {} ;
	\node[mycircle,below left=30pt and 0pt of sourcex] (middle2) {} ;
	\node[mycircle,below left=30pt and -20pt of sourcex] (middle22) {} ;
	\node[rectangle,right = 20pt of middle2](text1){$\dots$};
	\node[mycircle,below left=30pt and -60pt of sourcex] (middle3) {} ;
	\node[mycircle,below left=30pt and -80pt of sourcex] (middle4) {} ;
	\node[rectangle,right = 5pt of middle4](text2){$r$ nodes};

%	\draw[black,->,very thick] (sourcex.south) -- (middle0.north);
%	\draw[black,->,very thick] (sourcex.south) -- (middle1.north);
%	\draw[black,->,very thick] (sourcex.south) -- (middle2.north);
%	\draw[black,->,very thick] (sourcex.south) -- (middle3.north);
%	\draw[black,->,very thick] (sourcex.south) -- (middle4.north);

	\node[mycircle,below left=30pt and 10pt of middle0] (rec1) {} ;
	\node[mycircle,below left=30pt and -20pt of middle0] (rec2) {} ;
	\node[mycircle,below left=30pt and -150pt of middle0] (reclast) {} ; %,label=below:{R$\binom{r}{s}$}] (reclast) {} ;
	
	\draw[black,->,very thick] (middle2.south) -- (rec1.north);
	\draw[black,->,very thick] (middle22.south) -- (rec2.north);
	\draw[black,->,very thick] (middle1.south) -- (rec1.north);
%	\draw[black,->,very thick] (middle0.south) -- (rec2.north);
	\draw[black,->,very thick] (middle2.south) -- (rec2.north);
	\draw[black,->,very thick] (middle3.south) -- (reclast.north);
	\draw[black,->,very thick] (middle4.south) -- (reclast.north);
	\node[rectangle,right = 23pt of middle2](blub){$\dots$};
	\draw[black,->,very thick] (blub.south) -- (reclast.north);
	\node[rectangle,below right = 5pt and 10pt of middle4](text2){$s=4$ {incoming}};
	\node[rectangle,below right = 15pt and 26pt of middle4](text2){edges};
	\node[rectangle,below right = 30pt and 22pt of middle4](text2){$\binom{r}{3}$ receivers};
	
		\draw [-, very thick,black] (2.1,-2.8) arc (170:80:20pt);
	
\end{tikzpicture}

%% file: comb-network-rate2-ell.tex
\centering
\tikzsetnextfilename{2-butterflies}

%\tikzset{EdgeStyle/.append style = {->, bend left} }
  
\begin{tikzpicture}[scale = 1]
	% one butterfly
	\node[mycircle,label=right:{$\vec{x}_1,\dots,\vec{x}_h$}] (sourcex) {} 
	edge[bend left=5, very thick,->] (middle0.north)
	edge[bend right=5, very thick,->] (middle0.north)
	edge[bend left=5, very thick,->] (middle1.north)
	edge[bend right=5, very thick,->] (middle1.north)
	edge[bend left=5, very thick,->] (middle2.north)
	edge[bend right=5, very thick,->] (middle2.north)
	edge[bend left=5, very thick,->] (middle3.north)
	edge[bend right=5, very thick,->] (middle3.north)
	edge[bend left=5, very thick,->] (middle4.north)
	edge[bend right=5, very thick,->] (middle4.north)
	
	edge[bend right=40, very thick,->] (rec1.west)
	edge[bend left=40, very thick,->] (rec2.east)
	edge[bend right=20, very thick,->] (reclast.west)
	; 
	
	\draw [-, very thick,black] (1.7,-1) arc (-110:-40:10pt);
	\draw [-, very thick,black] (0.9,-.85) arc (-110:-40:10pt);
	\draw [-, very thick,black] (-1.45,-.7) arc (-160:-100:10pt);
	\draw [-, very thick,black] (-0.9,-.7) arc (-140:-70:10pt);
	\draw [-, very thick,black] (-0.45,-.7) arc (-120:-50:10pt);
	
	\draw [-, very thick,black] (1.93,-2) arc (-110:-40:10pt);
	\draw [-, very thick,black] (2.4,-2) arc (-110:-40:10pt);
	\draw [-, very thick,black] (-1.2,-2) arc (-140:-60:10pt);
	\draw [-, very thick,black] (-2.2,-2.1) arc (-160:-90:10pt);
	\draw [-, very thick,black] (-1.8,-2.2) arc (-110:-60:10pt);
	\draw [-, very thick,black] (-2.5,-2) arc (-160:-90:10pt);

	\node[rectangle, below right=6.5pt and 47pt of sourcex](textell){\textcolor{black}{$\boldsymbol{\ell}$}};
	
	\node[mycircle,below left=30pt and 40pt of sourcex] (middle0) {} 
	edge[bend left=5, very thick,->] (rec1.north)
	edge[bend right=5, very thick,->] (rec1.north)
	edge[bend left=5, very thick,->] (rec2.north)
	edge[bend right=5, very thick,->] (rec2.north)
	
	; 
	\node[mycircle,below left=30pt and 20pt of sourcex] (middle1) {} 
	edge[bend left=5, very thick,->] (rec1.north)
	edge[bend right=5, very thick,->] (rec1.north);
	\node[mycircle,below left=30pt and 0pt of sourcex] (middle2) {} 
	edge[bend left=5, very thick,->] (rec2.north)
	edge[bend right=5, very thick,->] (rec2.north);
	\node[rectangle,right = 10pt of middle2](text1){$\dots$}
	;
	\node[mycircle,below left=30pt and -60pt of sourcex] (middle3) {} 
	edge[bend left=5, very thick,->] (reclast.north)
	edge[bend right=5, very thick,->] (reclast.north);
	\node[mycircle,below left=30pt and -80pt of sourcex] (middle4) {}
	edge[bend left=5, very thick,->] (reclast.north)
	edge[bend right=5, very thick,->] (reclast.north) ;
	\node[rectangle,right = 5pt of middle4](text2){$r$ nodes};

%	\draw[black,->,very thick] (sourcex.south) -- (middle0.north);
%	\draw[black,->,very thick] (sourcex.south) -- (middle1.north);
%	\draw[black,->,very thick] (sourcex.south) -- (middle2.north);
%	\draw[black,->,very thick] (sourcex.south) -- (middle3.north);
%	\draw[black,->,very thick] (sourcex.south) -- (middle4.north);
	
	\node[mycircle,below left=30pt and 10pt of middle0] (rec1) {} ;
	\node[mycircle,below left=30pt and -20pt of middle0] (rec2) {} ;
	\node[mycircle,below left=30pt and -150pt of middle0] (reclast) {} ; %,label=below:{R$\binom{r}{s}$}] (reclast) {} ;
	
%	\draw[black,->,very thick] (middle0.south) -- (rec1.north);
%	\draw[black,->,very thick] (middle1.south) -- (rec1.north);
%	\draw[black,->,very thick] (middle0.south) -- (rec2.north);
%	\draw[black,->,very thick] (middle2.south) -- (rec2.north);
%	\draw[black,->,very thick] (middle3.south) -- (reclast.north);
%	\draw[black,->,very thick] (middle4.south) -- (reclast.north);
%	\node[rectangle,below right = 7pt and 7pt of middle4](text2){$s$ edges};
	
	\draw [-, very thick,black] (2.1,-2.8) arc (170:60:20pt);
	
		\node[rectangle,below right = 5pt and 16pt of middle4](text2){$s=2\ell+1$};
		\node[rectangle,below right = 15pt and 22pt of middle4](text2){incoming edges};
		\node[rectangle,below right = 30pt and 22pt of middle4](text2){$\binom{r}{2}$ receivers};
\end{tikzpicture}

%% file: comb-network-rate2-ell-plus.tex
\centering
\tikzsetnextfilename{2-butterflies}

%\tikzset{EdgeStyle/.append style = {->, bend left} }
  
\begin{tikzpicture}[scale = 1]
	% one butterfly
	\node[mycircle,label=right:{$\vec{x}_1,\dots,\vec{x}_h$}] (sourcex) {} 
	edge[bend left=5, very thick,->] (middle0.north)
	edge[bend right=5, very thick,->] (middle0.north)
	edge[bend left=5, very thick,->] (middle1.north)
	edge[bend right=5, very thick,->] (middle1.north)
	edge[bend left=5, very thick,->] (middle2.north)
	edge[bend right=5, very thick,->] (middle2.north)
	edge[bend left=5, very thick,->] (middle3.north)
	edge[bend right=5, very thick,->] (middle3.north)
	edge[bend left=5, very thick,->] (middle4.north)
	edge[bend right=5, very thick,->] (middle4.north)
	
	edge[bend right=40, very thick,->] (rec1.west)
	edge[bend right=35, very thick,->] (rec1.west)
	edge[bend left=40, very thick,->] (rec2.east)
	edge[bend left=35, very thick,->] (rec2.east)
	edge[bend right=20, very thick,->] (reclast.west)
	edge[bend right=15, very thick,->] (reclast.west)
	; 
	
	\draw [-, very thick,black] (.25,-1) arc (-110:-40:10pt);
	\draw [-, very thick,black] (-.03,-1) arc (-100:-40:10pt);
	\draw [-, very thick,black] (-2.4,-.9) arc (-160:-60:10pt);
	
	\draw [-, very thick,black] (1.7,-1) arc (-110:-40:10pt);
	\draw [-, very thick,black] (0.9,-.85) arc (-110:-40:10pt);
	\draw [-, very thick,black] (-1.45,-.7) arc (-160:-100:10pt);
	\draw [-, very thick,black] (-0.9,-.7) arc (-140:-70:10pt);
	\draw [-, very thick,black] (-0.45,-.7) arc (-120:-50:10pt);
	
	\draw [-, very thick,black] (1.93,-2) arc (-110:-40:10pt);
	\draw [-, very thick,black] (2.4,-2) arc (-110:-40:10pt);
	\draw [-, very thick,black] (-1.2,-2) arc (-140:-60:10pt);
	\draw [-, very thick,black] (-2.2,-2.1) arc (-160:-90:10pt);
	\draw [-, very thick,black] (-1.8,-2.2) arc (-110:-60:10pt);
	\draw [-, very thick,black] (-2.5,-2) arc (-160:-90:10pt);

	\node[rectangle, below right=6.5pt and 47pt of sourcex](textell){\textcolor{black}{$\boldsymbol{\ell}$}};
	
	\node[rectangle, below left=6.5pt and 48pt of sourcex](textell2){\textcolor{black}{$\boldsymbol{\ell-1}$}};
	
	\node[mycircle,below left=30pt and 40pt of sourcex] (middle0) {} 
	edge[bend left=5, very thick,->] (rec1.north)
	edge[bend right=5, very thick,->] (rec1.north)
	edge[bend left=5, very thick,->] (rec2.north)
	edge[bend right=5, very thick,->] (rec2.north)
	
	; 
	\node[mycircle,below left=30pt and 20pt of sourcex] (middle1) {} 
	edge[bend left=5, very thick,->] (rec1.north)
	edge[bend right=5, very thick,->] (rec1.north);
	\node[mycircle,below left=30pt and 0pt of sourcex] (middle2) {} 
	edge[bend left=5, very thick,->] (rec2.north)
	edge[bend right=5, very thick,->] (rec2.north);
	\node[rectangle,right = 10pt of middle2](text1){$\dots$}
	;
	\node[mycircle,below left=30pt and -60pt of sourcex] (middle3) {} 
	edge[bend left=5, very thick,->] (reclast.north)
	edge[bend right=5, very thick,->] (reclast.north);
	\node[mycircle,below left=30pt and -80pt of sourcex] (middle4) {}
	edge[bend left=5, very thick,->] (reclast.north)
	edge[bend right=5, very thick,->] (reclast.north) ;
	\node[rectangle,right = 5pt of middle4](text2){$r$ nodes};

%	\draw[black,->,very thick] (sourcex.south) -- (middle0.north);
%	\draw[black,->,very thick] (sourcex.south) -- (middle1.north);
%	\draw[black,->,very thick] (sourcex.south) -- (middle2.north);
%	\draw[black,->,very thick] (sourcex.south) -- (middle3.north);
%	\draw[black,->,very thick] (sourcex.south) -- (middle4.north);
	
	\node[mycircle,below left=30pt and 10pt of middle0] (rec1) {} ;
	\node[mycircle,below left=30pt and -20pt of middle0] (rec2) {} ;
	\node[mycircle,below left=30pt and -150pt of middle0] (reclast) {} ; %,label=below:{R$\binom{r}{s}$}] (reclast) {} ;
	
%	\draw[black,->,very thick] (middle0.south) -- (rec1.north);
%	\draw[black,->,very thick] (middle1.south) -- (rec1.north);
%	\draw[black,->,very thick] (middle0.south) -- (rec2.north);
%	\draw[black,->,very thick] (middle2.south) -- (rec2.north);
%	\draw[black,->,very thick] (middle3.south) -- (reclast.north);
%	\draw[black,->,very thick] (middle4.south) -- (reclast.north);
%	\node[rectangle,below right = 7pt and 7pt of middle4](text2){$s$ edges};

	\draw [-, very thick,black] (2.1,-2.8) arc (170:60:20pt);
	
		\node[rectangle,below right = 5pt and 16pt of middle4](text2){$s=3\ell-1$};
		\node[rectangle,below right = 15pt and 23pt of middle4](text2){incoming edges};
		\node[rectangle,below right = 30pt and 22pt of middle4](text2){$\binom{r}{2}$ receivers};
	
\end{tikzpicture}

%% file: comb-network-rate2.tex
\centering
\tikzsetnextfilename{2-butterflies}

%\tikzset{EdgeStyle/.append style = {->, bend left} }
  
\begin{tikzpicture}[scale = 1]
	% one butterfly
	\node[mycircle,label=right:{$\vec{x}_1,\vec{x}_2,\vec{x}_3,\vec{x}_4$}] (sourcex) {} 
	edge[bend left=5, very thick,->] (middle0.north)
	edge[bend right=5, very thick,->] (middle0.north)
	edge[bend left=5, very thick,->] (middle1.north)
	edge[bend right=5, very thick,->] (middle1.north)
	edge[bend left=5, very thick,->] (middle2.north)
	edge[bend right=5, very thick,->] (middle2.north)
	edge[bend left=5, very thick,->] (middle3.north)
	edge[bend right=5, very thick,->] (middle3.north)
	edge[bend left=5, very thick,->] (middle4.north)
	edge[bend right=5, very thick,->] (middle4.north)
	
	edge[bend right=40, very thick,->] (rec1.west)
	edge[bend left=40, very thick,->] (rec2.east)
	edge[bend right=20, very thick,->] (reclast.west)
	; 
	
	\node[mycircle,below left=30pt and 40pt of sourcex] (middle0) {} 
	edge[bend left=5, very thick,->] (rec1.north)
	edge[bend right=5, very thick,->] (rec1.north)
	edge[bend left=5, very thick,->] (rec2.north)
	edge[bend right=5, very thick,->] (rec2.north)
	
	; 
	\node[mycircle,below left=30pt and 20pt of sourcex] (middle1) {} 
	edge[bend left=5, very thick,->] (rec1.north)
	edge[bend right=5, very thick,->] (rec1.north);
	\node[mycircle,below left=30pt and 0pt of sourcex] (middle2) {} 
	edge[bend left=5, very thick,->] (rec2.north)
	edge[bend right=5, very thick,->] (rec2.north);
	\node[rectangle,right = 10pt of middle2](text1){$\dots$}
	;
	\node[mycircle,below left=30pt and -60pt of sourcex] (middle3) {} 
	edge[bend left=5, very thick,->] (reclast.north)
	edge[bend right=5, very thick,->] (reclast.north);
	\node[mycircle,below left=30pt and -80pt of sourcex] (middle4) {}
	edge[bend left=5, very thick,->] (reclast.north)
	edge[bend right=5, very thick,->] (reclast.north) ;
	\node[rectangle,right = 5pt of middle4](text2){$r$ nodes};

%	\draw[black,->,very thick] (sourcex.south) -- (middle0.north);
%	\draw[black,->,very thick] (sourcex.south) -- (middle1.north);
%	\draw[black,->,very thick] (sourcex.south) -- (middle2.north);
%	\draw[black,->,very thick] (sourcex.south) -- (middle3.north);
%	\draw[black,->,very thick] (sourcex.south) -- (middle4.north);
	
	\node[mycircle,below left=30pt and 10pt of middle0] (rec1) {} ;
	\node[mycircle,below left=30pt and -20pt of middle0] (rec2) {} ;
	\node[mycircle,below left=30pt and -150pt of middle0] (reclast) {} ; %,label=below:{R$\binom{r}{s}$}] (reclast) {} ;
	
%	\draw[black,->,very thick] (middle0.south) -- (rec1.north);
%	\draw[black,->,very thick] (middle1.south) -- (rec1.north);
%	\draw[black,->,very thick] (middle0.south) -- (rec2.north);
%	\draw[black,->,very thick] (middle2.south) -- (rec2.north);
%	\draw[black,->,very thick] (middle3.south) -- (reclast.north);
%	\draw[black,->,very thick] (middle4.south) -- (reclast.north);
%	\node[rectangle,below right = 7pt and 7pt of middle4](text2){$s$ edges};
		\draw [-, very thick,black] (2.1,-2.8) arc (170:60:20pt);
		
			\node[rectangle,below right = 5pt and 16pt of middle4](text2){$s=5$};
			\node[rectangle,below right = 13pt and 22pt of middle4](text2){incoming edges};	
			\node[rectangle,below right = 30pt and 22pt of middle4](text2){$\binom{r}{2}$ receivers};
\end{tikzpicture}

%% file: VecNetCode-final-v2.bbl
\begin{thebibliography}{10}
\providecommand{\url}[1]{#1}
\csname url@samestyle\endcsname
\providecommand{\newblock}{\relax}
\providecommand{\bibinfo}[2]{#2}
\providecommand{\BIBentrySTDinterwordspacing}{\spaceskip=0pt\relax}
\providecommand{\BIBentryALTinterwordstretchfactor}{4}
\providecommand{\BIBentryALTinterwordspacing}{\spaceskip=\fontdimen2\font plus
\BIBentryALTinterwordstretchfactor\fontdimen3\font minus
  \fontdimen4\font\relax}
\providecommand{\BIBforeignlanguage}[2]{{%
\expandafter\ifx\csname l@#1\endcsname\relax
\typeout{** WARNING: IEEEtranS.bst: No hyphenation pattern has been}%
\typeout{** loaded for the language `#1'. Using the pattern for}%
\typeout{** the default language instead.}%
\else
\language=\csname l@#1\endcsname
\fi
#2}}
\providecommand{\BIBdecl}{\relax}
\BIBdecl

\bibitem{Ahlswede_NetworkInformationFlow_2000}
R.~Ahlswede, N.~Cai, S.-Y. R.~Li, and R. W.~Yeung, ``{Network information flow},''
\emph{IEEE Trans. Inform. Theory}, vol.~46, no.~4, pp. 1204--1216, Aug. 2000.

\bibitem{CannonsDoughertyFreilingZeger-2006}
J. Cannons, R. Dougherty, C. Freiling, K. Zeger, ``{Network routing capacity},
'' \emph{IEEE Trans. Inform. Theory}, vol.~52, no.~3, pp. 777--788, Mar. 2006.

\bibitem{ConnellyZeger-2016-2}
J.~Connelly and K.~Zeger, ``Linear network coding over rings part I: scalar codes and commutative alphabets'',
[Online:] https://arxiv.org/abs/1608.01738, 2016,
\emph{IEEE Trans. Inform. Theory}, to appear, 2017.

\bibitem{ConnellyZeger-2016-1}
J.~Connelly and K.~Zeger, ``Linear network coding over rings part II: vector codes and
non-commutative alphabets'', [Online:] https://arxiv.org/abs/1608.01737, 2016,
\emph{IEEE Trans. Inform. Theory}, to appear, 2017.

\bibitem{DasRai-2016}
N.~Das and B.~K.~Rai, ``On the message dimensions of vector linearly solvable networks'',
\emph{IEEE Comm. Letters}, vol.~20, no.~9, pp. 1701--1704, 2016.

\bibitem{Delsarte_1978}
P.~Delsarte, ``{Bilinear forms over a finite field with applications to coding
theory},'' \emph{J. Combin. Theory Ser. A}, vol.~25, no.~3, pp. 226--241,
1978.

\bibitem{Didier-ErasureRS}
F.~Didier, ``{Efficient erasure decoding of Reed--Solomon codes},'' in
\emph{IEEE Int. Symp. Inform. Theory (ISIT)}, 2009.

\bibitem{DoughertyFreilingZeger-InsufficiencyOfLinearCodingInNetworkInformationFlow_2005}
R.~Dougherty, C.~Freiling, and K.~Zeger,
``{Insufficiency of linear coding in network information flow},''
\emph{IEEE Trans. Inform. Theory}, vol.~51, no.~8,
pp. 2745--2759, Aug. 2005.

\bibitem{DoughertyFreilingZeger-LinearitySolvabilityMulticastNetworks_2004}
R.~Dougherty, C.~Freiling, and K.~Zeger, ``{Linearity and solvability in multicast networks},''
\emph{IEEE Trans. Inform. Theory}, vol.~50, no.~10,
pp. 2243--2256, Oct. 2004.

\bibitem{DoughertyFreilingZeger-NetworksMatroidsNonShannon_2007}
R.~Dougherty, C.~Freiling, and K.~Zeger, ``{Networks, matroids, and non-Shannon
information inequalities},'' \emph{IEEE Trans. Inform. Theory}, vol.~53, no.~6,
pp. 1949--1969, Jun. 2007.

\bibitem{EbrahimiFragouli-AlgebraicAlgosVectorNetworkCoding}
J.~B.~Ebrahimi and C.~Fragouli, ``{Algebraic algorithms for vector network
coding},'' \emph{IEEE Trans. Inform. Theory}, vol.~57, no.~2, pp. 996--1007,
Feb. 2011.

\bibitem{EtzionGorlaRavagnaniWachterzeh_Ferrers_2014}
T.~Etzion, E.~Gorla, A.~Ravagnani, and A.~{Wachter-Zeh}, ``{Optimal Ferrers
  Diagram Rank-Metric Codes},'' \emph{IEEE Trans. Inform. Theory}, vol.~62,
  no.~4, pp. 1616--1630, Apr. 2016.

\bibitem{EOO18}
T.~Etzion, K.~Otal, and F.~\"{O}zbudak,
    manuscipt in preparation.

\bibitem{Etzion2009ErrorCorrecting}
T.~Etzion and N.~Silberstein, ``{Error-correcting codes in projective spaces
via rank-metric codes and Ferrers diagrams},'' \emph{IEEE Trans. Inform.
Theory}, vol.~55, no.~7, pp. 2909--2919, Jul. 2009.

\bibitem{EtzionSilberstein-CodesDesignsRelLiftedMRD-2013}
------, ``{Codes and designs related to lifted MRD codes},'' \emph{IEEE Trans.
  Inform. Theory}, vol.~59, pp. 1004--1017, Feb. 2013.

\bibitem{Etzion_Storme_2015}
T.~Etzion and L.~Storme, ``{Galois geometries and coding theory},''
\emph{Designs, Codes and Cryptography}, vol.~78, no.~1, pp. 311--350, 2016.

\bibitem{Etzion2011ErrorCorrecting}
T.~Etzion and A.~Vardy, ``{Error-correcting codes in projective space},''
  \emph{IEEE Trans. Inform. Theory}, vol.~57, no.~2, pp. 1165--1173, Feb. 2011.

\bibitem{EtzionWachterzeh-ISIT2016}
T.~Etzion and A.~Wachter-Zeh, ``Vector network coding based on subspace codes outperforms scalar linear network coding,''
\emph{IEEE Int. Symp. Inform Theory (ISIT)}, pp. 1949--1953, Jul. 2016.

\bibitem{EtZh18}
T.~Etzion and H.~Zhang, ``Grassmannian codes with new distance measures for network coding'',
[Online:] https://arxiv.org/abs/1801.02329, 2018,

\bibitem{Fragouli-Sojanin-NetworkCodingMulticast_2015}
C.~Fragouli and E.~Soljanin, ``{(Secure) linear network coding multicast},''
\emph{Designs, Codes and Cryptography}, vol.~78, no.~1, pp. 269--310, 2016.

\bibitem{Gabidulin_TheoryOfCodes_1985}
E.~M. Gabidulin, ``{Theory of codes with maximum rank distance},'' \emph{Probl.
Inform. Transm.}, vol.~21, no.~1, pp. 3--16, Mar. 1985.

\bibitem{Gathen_Gerhard_computer_algebra}
J.~Gathen and J.~Gerhard, \emph{{Modern computer algebra}}.\hskip 1em plus
0.5em minus 0.4em\relax Cambridge University Press, 2003.

\bibitem{Gohberg94fastalgorithms}
I.~Gohberg and V.~Olshevsky, ``{Fast algorithms with preprocessing for
matrix-vector multiplication problems},'' \emph{Common Divisors of Matrix
Polynomials, I. Spectral Method”, Indiana Journal of Math}, vol.~30, pp.
1981--321, 1994.

\bibitem{JaggiCassutoEffros-LowComplexityVectorEncoding}
S.~Jaggi, Y.~Cassuto, and M.~Effros, ``{Low complexity encoding for network
  codes},'' in \emph{IEEE Inform. Symp. Inform. Theory (ISIT)},
  Jul. 2006, pp. 40--44.

\bibitem{JaggiSanders-PolyTimeAlgoMulticastNetworkCode}
S.~Jaggi, P.~Sanders, P.~Chou, M.~Effros, S.~Egner, K.~Jain, and L.~M.
Tolhuizen, ``{Polynomial time algorithms for multicast network code
construction},'' \emph{IEEE Trans. Inform. Theory}, vol.~51, no.~6, pp.
1973--1982, Jun. 2005.

\bibitem{koetter_kschischang}
R.~K\"otter and F.~R. Kschischang, ``{Coding for errors and erasures in random
  network coding},'' \emph{IEEE Trans. Inform. Theory}, vol.~54, no.~8, pp.
  3579--3591, Jul. 2008.


\bibitem{KoetterMedard-AlgebraicApproachNetworkCoding_Journal}
R.~K\"otter and M.~M\'{e}dard, ``{An algebraic approach to network coding},''
\emph{IEEE/ACM Trans. on Networking}, vol.~11, no.~5, pp. 782--795, Oct.
2003.

\bibitem{LangbergSprintsonBruck-NetworkCodingComputational}
M. Langberg, A. Sprintson, and J. Bruck,
``{Network coding: A computational perspective},'' \emph{IEEE Trans. Inform. Theory}, vol.~55, no.~1, pp.~147--157, Jan. 2009.

\bibitem{LangbergSprintson-RecentResultsComplexityNEC}
M. Langberg and A. Sprintson, ``{Recent results on the algorithmic complexity of network coding},''
\emph{Tutorial appearing in proceedings of 2009 Workshop on Network Coding, Theory, and Applications
(NetCod 2009)}, 2009.

\bibitem{LangbergSprintson-HardnessApproxmatingNetworkCodingCapacity}
M. Langberg and A. Sprintson, ``{On the hardness of approximating the network coding capacity},'' \emph{IEEE Int. Symp. Inform. Theory (ISIT)}, Toronto, Canada, 2008.

\bibitem{LehmanLehman-ComplexityNetworkInfoFlow}
A.~Lehman and E.~Lehman, ``{Complexity classification of network information
flow problems},'' in \emph{15th ACM-SIAM Symposium on Discrete Algorithms
(SODA)}, 2004, pp. 142--150.

\bibitem{LiYeungCai-LinearNetworkCoding-2003}
S.-Y.~R. Li, R.~W.~Yeung, and N. Cai, ``Linear network coding'', \emph{IEEE Trans. Inform. Theory}, vol.~49, no.~2, pp. 371--381, 2003.


\bibitem{Lusina2003Maximum}
P.~Lusina, E.~M. Gabidulin, and M.~Bossert, ``{Maximum rank distance codes as
space-time codes},'' \emph{IEEE Trans. Inform. Theory}, vol.~49, no.~10, pp.
2757--2760, Oct. 2003.

\bibitem{MacWilliamsSloane_TheTheoryOfErrorCorrecting_1988}
F.~J.~MacWilliams and N.~J.~A.~Sloane, \emph{{The theory of error-correcting
codes}}.\hskip 1em plus 0.5em minus 0.4em\relax North Holland Publishing Co.,
1988.

\bibitem{MedardKargerEffrosKargerHo_2003}
M. Medard, M. Effros, D. Karger, and T. Ho,``{On coding for non-multicast networks},'' \emph{In Proceedings of the 41st Allerton
Conference on Communication, Control and Computing}, Monticello, IL, October 2003.

\bibitem{Menezes2010Applications}
A.~J.~Menezes, I.~F.~Blake, X.~Gao, R.~C. Mullin, S.~A.~Vanstone, and T.~Yaghoobian, \emph{Applications of finite fields}.\hskip 1em plus 0.5em minus 0.4em\relax Springer, 1993.


\bibitem{RiisAhlswede-ProblemsNetworkCodingECC}
S.~Riis and R.~Ahlswede, ``{Problems in network coding and error correcting
codes},'' in \emph{General Theory of Information Transfer and Combinatorics},
ser. Lecture Notes in Computer Science.\hskip 1em plus 0.5em minus
0.4em\relax Springer Berlin Heidelberg, 2006, vol. 4123, pp. 861--897.

\bibitem{Roth_RankCodes_1991}
R.~M. Roth, ``{Maximum-rank array codes and their application to crisscross
error correction},'' \emph{IEEE Trans. Inform. Theory}, vol.~37, no.~2, pp.
328--336, Mar. 1991.

\bibitem{Roth16}
R.~M. Roth, \emph{Personal communication}, 2016.

\bibitem{silva_rank_metric_approach}
D.~Silva, F.~R. Kschischang, and R.~K{\"o}tter, ``{A rank-metric approach to error control in random network coding},'' \emph{IEEE Trans. Inform. Theory},
  vol.~54, no.~9, pp. 3951--3967, 2008.

\bibitem{SunYangLongLi-MulticastNetworksVectorLinearCoding}
Q.~Sun, X.~Yang, K.~Long, and Z.~Li, ``{Constructing multicast networks where
vector linear coding outperforms scalar linear coding},'' in \emph{IEEE Int.
Symp. Inform. Theory (ISIT)}, 2015.

\bibitem{SunYangLongLi-MulticastNetworksVectorLinearCoding-arxiv}
Q.~Sun, X.~Yang, K.~Long, X.~Yin, and Z.~Li,  ``{On vector linear solvability of multicast networks},'' \emph{IEEE Trans. Comm.}, vol.~64, no.~12, pp.~5096--5107, 2016.
%[Online:] http://arxiv.org/abs/1605.02635, 2016.

\bibitem{SunYinLiLong-MulticastNCandFielsSize}
Q.~Sun, X.~Yin, Z.~Li, and K.~Long, ``Multicast network coding and field
  size,'' in \emph{IEEE Int. Symp. Inform. Theory (ISIT)}, 2014.

\bibitem{Turner-InverseVandermonde}
L.~R. Turner, ``{Inverse of the Vandermonde matrix with applications},''
\emph{NASA Technical Note}, vol. D-3547, Aug. 1966.

%\bibitem{MuralidharanRajan-2016}
%V.~T.~Muralidharan, B.~S.~Rajan, ``Linear network coding, linear index coding and representable discrete polymatroids,'' \emph{IEEE Trans. Inform. Theory},
%vol~62, no.~7, pp. 4096--4119, 2016.


\end{thebibliography}
